%% file: UniTok-Audio-1029.tex
\PassOptionsToPackage{colorlinks=true, citecolor=blue, urlcolor=blue, linkcolor=black}{hyperref}
\documentclass{article} 
\usepackage{iclr2026_conference,times}

\input{math_commands.tex}

\usepackage{adjustbox}  
\usepackage{booktabs}
\usepackage{makecell}
\usepackage{caption}
\usepackage{graphicx}
\usepackage{hyperref}
\usepackage{booktabs}
\usepackage{amsmath}
\usepackage{amssymb}
\usepackage{url}
\usepackage{multirow}
\usepackage{siunitx}
\usepackage{cleveref}
\usepackage{pifont}

\usepackage{threeparttable}  
\usepackage[table]{xcolor}
\usepackage{subcaption}

\title{UniTok-Audio: A Unified Audio Generation Framework Via Universal Discrete Token}
\title{UniTok-Audio: A Unified Audio Generation Framework via Generative Modeling on Discrete Codec Tokens}

\iclrfinalcopy

\author{Chengwei Liu$^{1,}$\thanks{Equal contribution.} , Haoyin Yan$^{1,*}$, Shaofei Xue$^{1,2,}$\thanks{Corresponding Author.} , Xiaotao Liang$^{1}$, Yinghao Liu$^1$\\ \textbf{Zheng Xue$^1$, Gang Song$^1$,Boyang Zhou$^{1,3}$}\\
	$^1$Intelligent Connectivity, Alibaba Group, $^2$TongYi Ai lab, Alibaba Group\\
	$^3$Zhejiang University\\
	\texttt{liuchengwei.lcw@alibaba-inc.com, shaofei.xsf@alibaba-inc.com} \\
}

\begin{document}

\maketitle

\begin{abstract}
Generative modeling has recently achieved remarkable success across text, image, and audio domains, 
demonstrating powerful capabilities for unified representation learning. 
However, audio generation models still face challenges in terms of audio quality and generalization ability across tasks. 
This fragmentation results in redundant development efforts, inconsistent performance, and limited extensibility.
To address these issues, we propose \textbf{UniTok-Audio}, a scalable and extensible framework for unified audio generation tasks.
Specifically, 1) UniTok-Audio extracts continuous feature of conditions to generates discrete tokens of target audio in an autoregressive manner; 
2) a special task identifier token unifies different learning patterns of multiple tasks in a single framework; 
3) a dual-stream audio codec involving acoustic and semantic branch is developed for high-fidelity waveform reconstruction. 
Experimental results demonstrate that UniTok-Audio achieves competitive performance 
in comparation with state-of-the-art task-specific or multi-task systems across five time-aligned tasks: 
speech restoration, target speaker extraction, speech separation, voice conversion, and language-queried audio source separation. 
To foster future research, we will open-source our codebase. 
The demo page of our work can be found here: https://alibaba.github.io/unified-audio.

\end{abstract}

\section{Introduction}
\label{sec:intro}

Leveraging the remarkable sequential generation capability of language model (LM)~\citep{vaswani2017attention}, 
recent works have achieved significant improvements in generation quality~\citep{polyak2024movie,lipman2023flow}, 
promoting the growing prevalence of artificial intelligence-generated content (AIGC). 
These advances have inspired substantial research extending LMs to various audio tasks, 
which can be fundamentally categorized by the temporal relationship between input and output: 
either \textit{time-aligned} (TA) or \textit{non-time-aligned} (NTA)~\citep{xu2025uniflowaudio}. 
The former involves strict temporal correspondence between input and output signals, such as speech denoising, 
which aligns speech components in each frame between noisy and clean speech. 
While the latter dose not require point-wise temporal alignment, such as text-to-audio (TTA), 
which aims at semantic coherence between the holistical textual description and entire output soundscape. 

This study focuses on the TA tasks, especially which provides the input audio that temporally aligned with the output audio at the frame level, 
including: speech restoration (SR) that aims at restoring speech from the degraded recording with various distortions (e.g., noise, reverberation,and packet loss); 
target speaker extraction (TSE) that extracts target speech from mixture using assistive clues (e.g., voiceprint information from reference speech); 
speech separation (SS) that aims to separate all existing speaker in the mixture; 
voice conversion (VC) that transforms the timbre of source speech guided by reference speech of another speaker; 
language-queried audio source separation (LASS) that aims at extracting target audio components from mixture, which are consistent with the given textual caption. 
Numerous generative models are developed for these tasks, while most of them are designed for single task with 
task-specific architectures~\citep{yuan2025flowsep, lee2025flowse, wang2024selm, LauraTSE, wang2023lm}. 
This fragmentation results in redundant development efforts, inconsistent performance, and limited extensibility. 

Some studies aim to unify multiple tasks within a single framework, including AnyEnhance~\citep{zhang2025anyenhance}, 
UniAudio~\citep{yang2024uniaudio}, LLaSE-G1~\citep{kang2025llaseg1}, UniSE~\citep{unise}, and Metis~\citep{wang2025metis}. 
These methods utilizes the LM backbone combined with discrete audio codec and 
exhibit remarkable generative ability, which benefit from the semantic understanding and contextual modeling capabilities of LMs.
However, challenges still exist in terms of audio quality and generalization ability across tasks. 
For instance, few unified models are capable of handling the SS task, as it generally requires customized 
architecture to output multi-track speech. 

To improve audio generation quality, some works~\citep{voicebox, audiobox, wang2025uniflow, xu2025uniflowaudio} 
adopt generative paradigms in continuous space, such as flow matching~\citep{lipman2023flow}, 
which eliminates the dependence on discrete codecs. 
However, the flowchart of model needs to be carefully designed to support different tasks, increasing the 
difficulty when extending to more tasks. 
Additionally, considering the trend of combining audio generation capabilities with large language models (LLM)~\citep{qwen3omni}, 
developing audio generation models based on discrete codec has greater potential. 
This highlights the need for improving the ability of audio codec, 
which directly affects the generation quality of audio models.

In this work, we propose \textbf{UniTok-Audio}, 
a novel decoder-only autoregressive (AR) LM-based generative framework to unify multiple TA tasks.
The contributions of this work can be summarized as follows:
\begin{enumerate}
	\item \textbf{Unified Framework}: The framework unifies tasks by taking task-specific conditional information as the conditioning sequence of decoder-only LM, 
	and the discrete token of target audio is predicted in an AR manner. 
	We utilize a special task token to distinguish different learning patterns of multiple tasks. 
	Note that our model handles diverse tasks using a single set of shared weights, thereby eliminating the need for task-specific weight adaptation.
	
	\item \textbf{New Tokenization}: We present \textbf{H-codec}, 
	which integrates self-supervised learning (SSL) representation within the audio tokenization and reconstruction process. 
	The features from waveform and SSL model are individually quantized, resulting dual-stream (acoustic and semantic) codec tokens. 
	H-Codec achieves remarkable audio reconstruction quality with a low frame rate, 
	improving both the efficiency and performance of downstream audio generation.
	
	\item \textbf{High-Fidelity Generation}: UniTok-Audio achieves high-fidelity generation quality 
	in terms of SR, TSE, SS, VC, and LASS tasks, demonstrating strong competitiveness compared to 
	state-of-the-art (SOTA) task-specific or multi-task baselines.
	
\end{enumerate}

\section{Related Work}
\subsection{Generative Modeling for Audio Tasks}
In the domain of TA audio tasks, early researches focus on discriminative modeling, 
which directly learns the mapping between input signal and target audio~\citep{tfmask,ConvTasNet}. 
However, the lack of generative ability limits their generalization in unseen scenarios and 
the performance in extreme situations~\citep{welker22_interspeech, over_suppression}. 
Many studies integrate generative modeling into audio tasks in recent years. 
For the SR task, SELM~\citep{wang2024selm} applies k-means to quantize noisy speech representations obtained  
by WavLM~\citep{chen2022wavlm} into discrete tokens, 
and then a Transformer-based speech LM maps the noisy tokens to clean tokens.
For the LASS task, FlowSep~\citep{yuan2025flowsep} 
learns linear flow trajectories from noise to target source features within the variational autoencoder (VAE) latent space, 
which are guided by the encoded text embeddings and the mixture audio. 
However, these models are designed for specific task, facing limited extensibility when migrating to more tasks. 

Creating an unified framework that can tackle diverse tasks 
stands as a critical research goal in the field of artificial intelligence. 
In the unification of audio tasks, the approaches can be divided into two categories: 
discrete audio codec based method and continuous representation based method. 
The former is based on the pre-trained audio codec, which encodes the waveform into discrete space and reconstructs 
audio signal from it. 
The generative ability of AR modeling or masked generative modeling~\citep{MaskGIT} is leveraged to 
generate discrete tokens of the target audio. 
For instance, UniAudio~\citep{yang2024uniaudio} tokenizes the target audio along with other condition modalities 
and then concatenates source-target pair as a single sequence, performing next-token prediction using LLM. 
Metis~\citep{wang2025metis} adopts a two-stage generation framework using masked generative modeling, 
which first generates SSL tokens and then predicts acoustic tokens based the former. 
Continuous representation based methods typically adopt diffusion~\citep{ho2020denoising} or flow matching techniques, 
eliminating the inevitable quantitative loss in discrete codec. 
VoiceBox~\citep{voicebox} performs flow matching on mel-spectrograms to unify tasks such as text-to-speech (TTS) and speech editing. 
UniFlow~\citep{wang2025uniflow} utilizes VAE to learn a compact latent representation of raw audio, 
coupled with a diffusion Transformer (DiT)~\citep{dit} that predicts latent updates. 

Compared to discrete audio codec based method, especially decoder-only AR models which can elegantly 
integrate conditional information as a prefix sequence, continuous methods usually requires complex design 
to combines multimodal conditions, limiting the extensibility to more tasks. 
In addition, discrete audio representation plays an important role in combining with LLM~\citep{qwen3omni}, 
bridging the natural language instructions and continuous waveform. 
Therefore, we develop a decoder-only AR LM-based framework (UniTok-Audio) to unify audio tasks. 
It utilizes continuous conditional embeddings to maximize the preservation of semantic and acoustic information, 
predicting multi-layer codec tokens which reduce the quantization loss.

\subsection{Neural Audio Codec}

Neural audio codecs utilize neural networks to obtain highly compressed discrete representations of audio waveforms and 
aim to reconstruct high-fidelity signal form discrete tokens. 
For instance, SoundStream~\citep{soundstream} utilizes residual vector quantization (RVQ) where 
each quantizer refines the residuals left by the previous one, 
obtaining parallel multi-layer tokens and achieving remarkable reconstruction quality. 
Many works including Encodec~\citep{encodec} and DAC~\citep{dac} follow this paradigm to improve performance. 

With the development of LM, the research focus of codecs has gradually shifted 
from reducing data transmission costs toward the integration with LM, which ensures the high quality of generated audio. 
This requires codecs~\citep{semanticodec,moshi} to preserve more semantic information that can be understood and modeled by LM. 
X-Codec~\citep{xcodec} integrates the representations from the pre-trained SSL model to enhance semantic preservation, 
improving both reconstruction quality and downstream TTS performance. 
Some studies~\citep{wavtokenizer,unicodec} explore single-layer codecs that are more suitable for autoregressive modeling in LM. 
X-Codec2~\citep{xcodec2} utilizes finite scalar quantization (FSQ)~\citep{fsq} to perform single-layer quantization, 
enlarging the code space. BiCodec~\citep{SparkTTS} generates a hybrid token stream combining semantic and global tokens, 
which are derived from a SSL model and a speaker verification model, respectively. 
However, single-layer codecs with a low frame rate still faces challenges in high-fidelity reconstruction~\citep{xcodec2}, e.g., speaker similarity. 

In practice, downstream LMs are capable of generating multi-layer tokens in parallel~\citep{musicgen, t5tts}, 
thereby relaxing the requirement for single-layer quantization. 
This paradigm relies more heavily on the modeling capacity of LMs, raising the upper bound of the codec's reconstruction capability.
In this context, the frame rate of codecs plays a critical role, which determines the number of time steps for inference. 
Our H-Codec benefits from the RVQ technique and SSL representations, achieving significant reconstruction quality in the domain of speech, music, and general audio. 
The low frame rate ensures efficient generation when integrated with our UniTok-Audio framework.

\section{UniTok-Audio}
\begin{figure*}[t]
	\centering
	\includegraphics[width=1.0\textwidth]{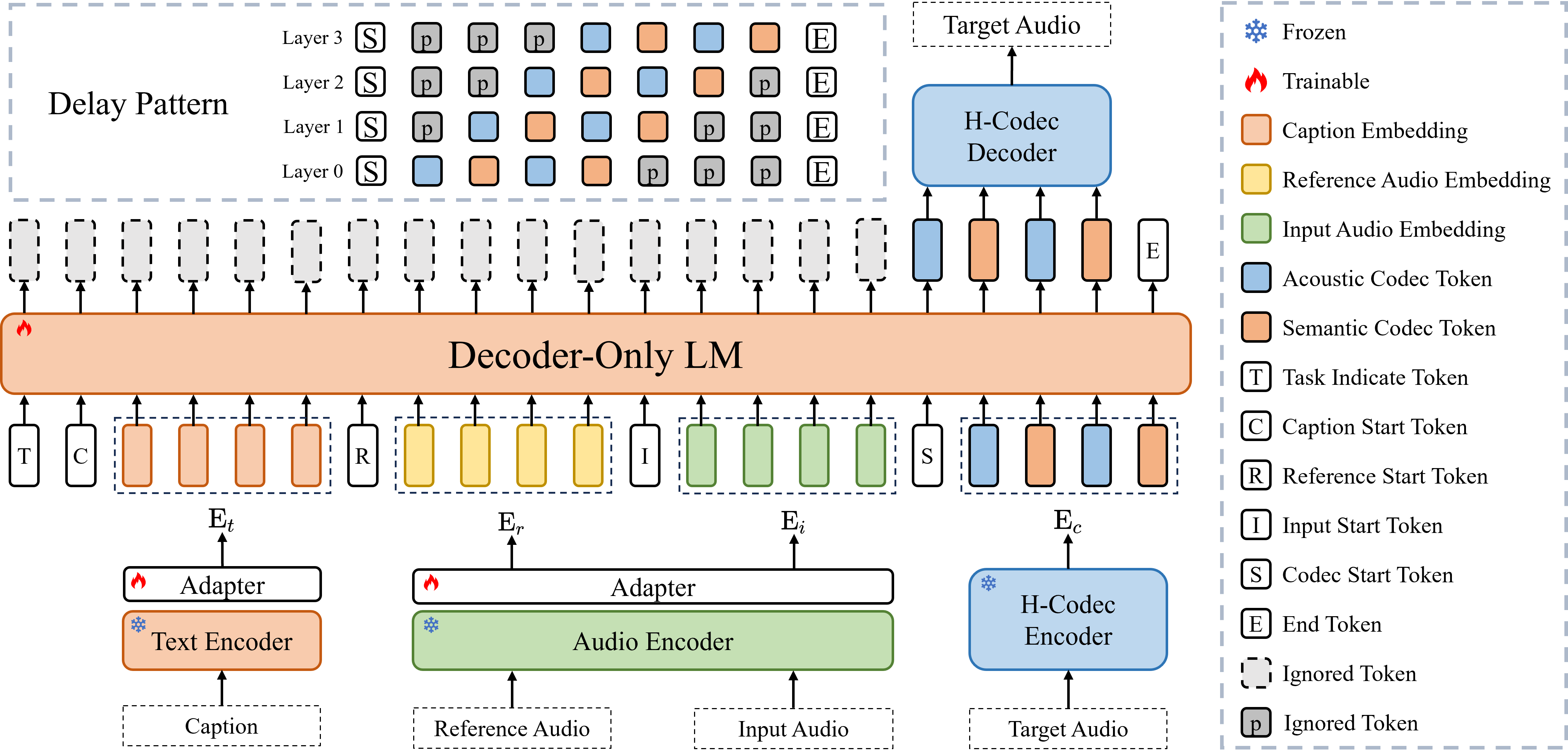}
	\caption{The overall architecture of UniTok-Audio, which is a straightforward model for multiple audio tasks. For simplicity, we illustrate the AR process with single-layer codec tokens and it actually operates in a multi-layer AR manner with delay pattern.}
	\label{fig:framework}
\end{figure*}

As shown in \Cref{fig:framework}, UniTok-Audio is a unified, autoregressive LM-based audio generation framework comprising four key components: 
(i) a novel dual-stream H-codec; (ii) a text encoder with adapter; (iii) an audio encoder with adapter; (iv) a decoder-only LM backbone.
Next, we will introduce the architecture of H-Codec and the operational framework of UniTok-Audio in detail.

\subsection{H-Codec}
\begin{figure*}
	\centering
	\includegraphics[width=\linewidth]{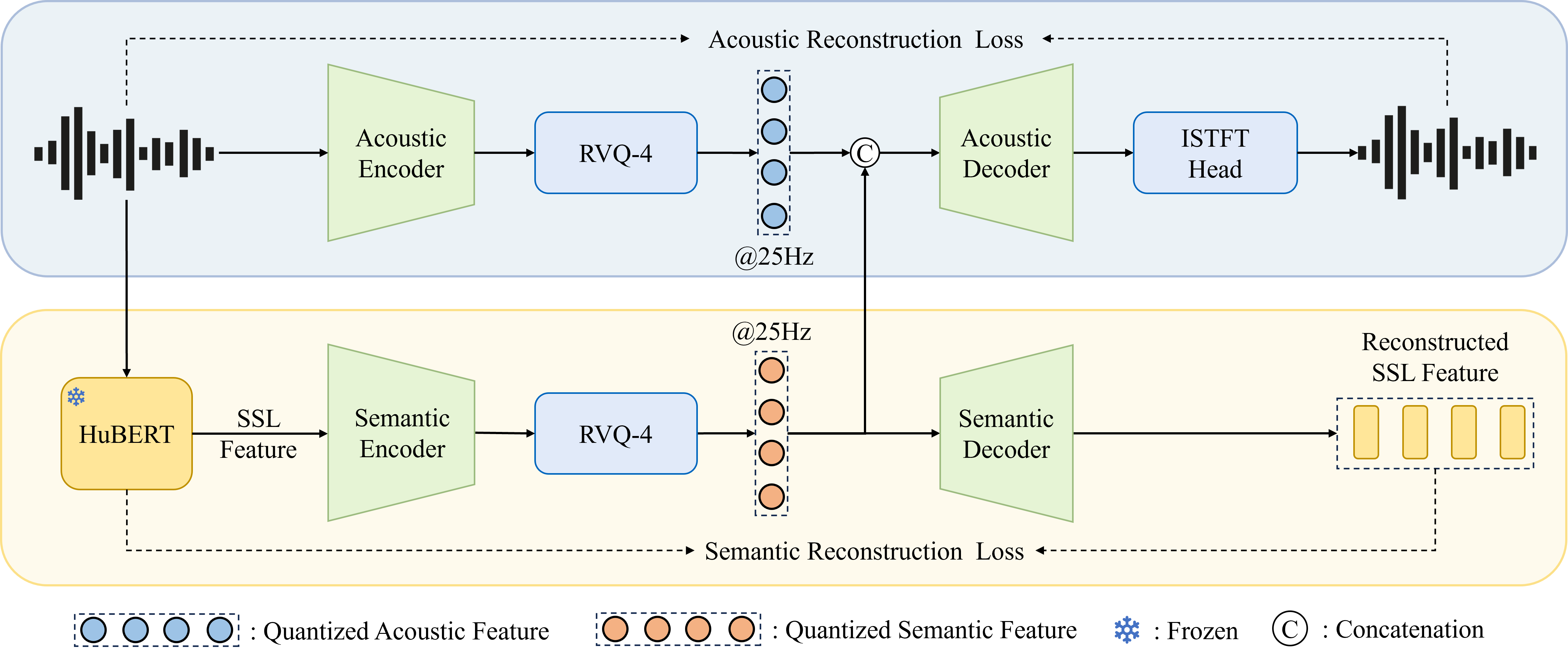}
	\caption{The framework of our proposed H-codec.}
	\label{fig:H-codec}    
\end{figure*}

To improve the audio generation quality, we propose H-Codec to discretize audio and reconstruct waveform from discrete tokens.
As illustrated in \Cref{fig:H-codec}, the architecture of H-Codec follows the common paradigm of audio tokenizers, including an acoustic encoder, a quantizer module, and an acoustic decoder.
Inspired by X-Codec~\citep{xcodec}, we incorporate pretrained models to facilitate the preservation of semantic information. 
However, unlike X-Codec, which fuses acoustic and semantic information and then quantizes the combined representation using a single codebook, 
we employ separate codebooks to quantize the two types of features independently, leading to dual-stream codec tokens.

\subsubsection{H-Codec Encoder}
In the encoding stage, the raw waveform $\boldsymbol{x} \in \mathbb{R}^{n}$ is fed into the acoustic encoder to extract frame-level acoustic features, where $n$ represents the number of waveform samples. 
The architecture of the acoustic encoder follows Encodec~\citep{encodec}. 
A 4-layer RVQ~\citep{soundstream} is utilized to quantize features, resulting in the quantized features with a frame rate of 25 Hz.
Synchronously, a pre-trained HuBERT\footnote{https://huggingface.co/bosonai/hubert\_base}~\citep{hubert} extracts SSL features by averaging outputs from all transformer layers
and the quantized semantic feature is obtained by applying the semantic encoder and RVQ quantizer. 
Note that HuBERT is trained on general audio, thus the codec has the potential to handle general audio as well.

\subsubsection{H-Codec Decoder}
For the waveform reconstruction, the quantized acoustic and semantic features are concatenated along the hidden dimension, and the 
waveform is reconstructed by utilizing acoustic decoder and the inverse short-time Fourier transform (ISTFT) head following Vocos~\citep{vocos}.
We believe that decoupling acoustic and semantic features enables each branch to learn distinct representations, which is beneficial for improving reconstruction quality.
Additionally, the quantized semantic feature is processed by the semantic decoder to reconstruct the SSL feature. 
This ensures that the quantized semantic features retain sufficiently rich semantic information. 

\subsubsection{Optimization Strategy}
The types of discriminators and the composition of the loss functions follow the configuration used in WavTokenizer~\citep{wavtokenizer}.
We employ a multi-period discriminator (MPD)~\citep{hifigan}, a multi-resolution discriminator (MRD)~\citep{univnet}, and a sub-band complex STFT discriminator~\citep{soundstream} to improve the naturalness and fidelity of reconstructed audio, 
and the training loss $\mathcal{L}_{dis}$ conforms to the hinge loss formulation suggested by~\citep{soundstream}. The training loss for the generator of H-Codec include: 
commitment loss for quantizer $\mathcal{L}_{commit}$, mel-spectrum reconstruction loss $\mathcal{L}_{mel}$, adversarial loss $\mathcal{L}_{adv}$, feature matching loss $\mathcal{L}_{fm}$, and an auxiliary mean squared error (MSE) loss on SSL feature $\mathcal{L}_{aux}$.
The composite training loss of the generator is obtained by
\begin{equation}
	\mathcal{L}_{gen} = \lambda_{commit} \mathcal{L}_{commit} + \lambda_{mel} \mathcal{L}_{mel} + \lambda_{adv} \mathcal{L}_{adv} + \lambda_{fm} \mathcal{L}_{fm} + \lambda_{aux} \mathcal{L}_{aux},
\end{equation}
where $\lambda_{commit}$, $\lambda_{mel}$, $\lambda_{adv}$, $\lambda_{fm}$, and $\lambda_{aux}$ are hyper-parameters to scale different loss components. 
Additionally, the perceptual loss~\citep{perceptual_loss} is utilized during the final steps of the training process, 
which further enhances the reconstruction quality.

\subsection{Unified Multi-Task Generation}

\subsubsection{Overall Framework}
To unify various audio generation tasks within a single framework, 
we extract task-specific conditional information as a conditioning sequence for the decoder-only AR backbone, 
which then predicts the corresponding H-Codec tokens of the target audio.
Since continuous features, typically extracted from SSL models, contain richer audio details compared to discrete representations and are more adaptable to varying input conditions, 
we extract continuous features to assemble the the task-conditioning sequence.
Specifically, we utilize T5-base\footnote{https://huggingface.co/google/t5-v1\_1-base}~\citep{t5} as the 
text encoder to extract embedding from audio caption.
The same HuBERT used in H-Codec is adopted to extract continuous features from audio waveforms.
Two linear layers serve as two adapters to map the text embedding and audio features into a representation space amenable to LM AR modeling, respectively.
Given text and audio embeddings as conditions, 
we utilize LLaMA architecture~\citep{LLaMA} to predicts discrete tokens of target waveform in an AR manner.
Finally, the H-Codec decoder reconstructs high-fidelity audio from the predicted token sequence.

\subsubsection{AR Prediction of H-Codec Tokens}
To incorporate multi-layer codec tokens into AR prediction, an existing method~\citep{valle} applies two-stage strategy: 
(i) model the tokens of the first layer in an AR manner; (ii) then, predict the tokens of remaining layers using a NAR post-network.
However, this method causes additional complexity to the system. In addition, flattening all tokens into one layer leads to unbearable computational 
cost, while predicting tokens from all layers in parallel within one step deteriorates the performance.
Therefore, we adopt the delay pattern~\citep{musicgen} to arrange our tokens for the trade-off between performance and computational cost.
Specifically, the 4-layer acoustic and semantic tokens produced by H-Codec are first interleaved sequentially across time steps, resulting in ${\rm E}_c \in \mathbb{Z}^{T \times 4}$
with a frame rate of 50 Hz, where $T$ indicates the number of frames. Before feeding the tokens into the LM backbone, 
different shifts are applied across layers and special pad tokens occupy empty positions, as shown in~\Cref{fig:framework}.
In the LM backbone, 4 embedding layers handle 4-layer tokens respectively, and the embeddings of each layer are added up as the input of transformer layers.
There are 4 output heads to predict the 4-layer logits of next time step. The delay pattern allows generating high-layer tokens conditioned by low-layer tokens, 
which improves prediction accuracy.

\begin{table}[tt]
	\centering
	\caption{Operational modes and corresponding conditions in UniTok-Audio.}
	\label{tab:conditioning}
	\small
	\begin{tabular}{@{}lcc@{}} 
		\toprule
		\textbf{Mode} & \textbf{Task Token} & \textbf{Conditions} \\
		\midrule
		\midrule
		SR & ${\rm T_{SR}}$ & Degraded Speech \\
		TSE & ${\rm T_{TSE}}$ & Reference Speech, Mixture Speech \\
		rTSE & ${\rm T_{rTSE}}$ & Reference Speech, Mixture Speech \\
		VC & ${\rm T_{VC}}$ & Reference Speech, Source Speech \\
		LASS & ${\rm T_{LASS}}$ & Caption, Mixture Audio \\
		\bottomrule
	\end{tabular}
\end{table}

\subsubsection{Unifying Tasks with Operational Modes}

Following our previous work~\citep{unise}, we introduce special task tokens to distinguish between different operational modes. 
To unify five tasks (i.e., SR, TSE, SS, VC, and LASS), we utilize five modes, as shown in Table~\ref{tab:conditioning}. 
Each mode corresponds to a special token and different task-specific condition types, which serve as a conditioning sequence 
for the LM backbone to estimate the conditional probability density distribution of target discrete tokens.

\textbf{SR Mode:} 
The target audio is the clean speech corresponding to the degraded input speech. The conditional sequence of LM is formatted as 
$\left[ {\rm T_{SR}}, {\rm I}, {\rm E}_i, {\rm S}\right]$, where ${\rm I}$ denotes the start of input audio features, 
${\rm E}_i$ the input audio embeddings, and ${\rm S}$ the start of codec tokens, respectively.
The output sequence is formulated as ${\bm o}= \left[ {\rm E}'_c, {\rm E} \right]$, where ${\rm E}'_c$ indicates codec tokens with 
delay pattern, and ${\rm E}$ represents the end token. The trainable parameters $\theta$ in the model 
are optimized by minimizing the negative log-likelihood of the predicted outputs:
\begin{align}
	\mathcal{L}_{\rm SR} = - \sum_{t=1}^{L} \sum_{i=1}^{4} {\rm log}P\left(o^i_t | {\rm T_{SR}}, {\rm I}, {\rm E}_i, {\rm S}, o_{<t} ; \theta\right),
\end{align}
where $o^i_t$ indicates the output token at $t$-th step and $i$-th layer, and $L$ is the length of output sequence, respectively.

\textbf{TSE Mode:}
The target audio corresponds to the timbre-matched speech component in the input mixture audio that aligns with the reference audio.
The conditional sequence is formatted as $\left[ {\rm T_{TSE}}, {\rm R}, {\rm E}_r, {\rm I}, {\rm E}_i, {\rm S}\right]$, 
where ${\rm E}_r$ and ${\rm R}$ represent the features of reference speech and its start token, respectively.
Therefore, the associated loss function is defined as 
\begin{align}
	\mathcal{L}_{\rm TSE} = - \sum_{t=1}^{L} \sum_{i=1}^{4} {\rm log}P\left(o^i_t | {\rm T_{TSE}}, {\rm R}, {\rm E}_r, {\rm I}, {\rm E}_i, {\rm S}, o_{<t} ; \theta\right).
	\label{loss_tse}
\end{align}

\textbf{rTSE Mode:} 
Since SS task requires generating multiple output tracks while our model only supports one-track output, 
we include the rTSE mode during training, 
enabling the model to obtain multiple tracks through iterative inference.
This mode aims to extract the timbre-mismatched speech component in the mixture input when compared with 
the reference speech. The loss function $\mathcal{L}_{\rm rTSE}$ keeps similar to that of the TSE mode, 
except that the task token has been replaced with ${\rm T_{rTSE}}$.
When handling SS task (we only consider 2-speaker cases), we first apply the SR mode to extract the main speaker 
with higher energy, and the other speaker is obtained by using the rTSE mode.

\textbf{VC Mode:} 
The target signal is timbre-perturbed version of input source speech using the speaker characteristics
of the reference speech, where the speech content remains unchanged.
The optimization object has a similar formulation with \eqref{loss_tse}.

\textbf{LASS Mode:} 
This mode aims at separating specific component that matches the given caption query from the input mixture audio.
Therefore, the associated  loss function is defined as
\begin{align}
	\mathcal{L}_{\rm LASS} = - \sum_{t=1}^{L} \sum_{i=1}^{4} {\rm log}P\left(o^i_t | {\rm T_{LASS}}, {\rm C}, {\rm E}_t, {\rm I}, {\rm E}_i, {\rm S}, o_{<t} ; \theta\right),
\end{align}
where ${\rm E}_t$ and ${\rm C}$ denote the embedding of caption and its start token, respectively.

\section{Experiments}

\subsection{H-Codec}
\subsubsection{Experimental Setup}
\textbf{Datasets:} 
We utilize multi-domain data to train our codec, including speech, music, and audio.
The speech samples are sourced from the VoxBox dataset~\citep{SparkTTS}, which comprises approximately 100k hours of speech 
and is composed of some publicly available speech datasets.
For the music domain, we utilize the FMA-full dataset~\citep{Defferrard2017FMA} and the MUSDB18-HQ dataset~\citep{MUSDB18HQ}, 
involving about 8k hours of data.
For the audio domain, we adopt AudioSet~\citep{gemmeke2017audio} and WavCaps~\citep{wavcaps}, 
including about 13k hours of recordings. 
We evaluate the reconstruction quality on LibriSpeech~\citep{Librispeech} test-clean, MUSDB18-HQ test, and AudioSet eval sets for 
speech, music, and audio domain, respectively.
All samples are resampled to 16k Hz. 

\textbf{Implementation Details:} 
The total downsampling ratio in H-Codec is set to 640 to obtain the frame rate of 25 Hz in both acoustic and semantic branch.
In the 4-layer RVQ, we utilize a codebook size of 1024 for each layer with the codebook dimension set to 512.
During training, we randomly crop 5-second segments from audio samples. 
The network is optimized using the AdamW optimizer with an initial learning rate of 
$2 \times 10^{-4}$, which is decayed based on a cosine scheduler.
In total, we train for 600k steps, and the perceptual loss is activated at final 100k steps.

\textbf{Evaluation Metrics:} 
We utilize several metrics to measure the reconstruction quality of speech, 
including the perceptual evaluation of speech quality (PESQ), short-time objective intelligibility (STOI), 
speaker similarity (SPK-SIM) and UTMOS.
The loss on Mel-scale spectrum and STFT spectrum bettween the target audio and reconstructed audio 
are computed for general evaluation in the domain of speech, music, and audio.
Details about evaluation metrics of codec can be found in Appendix~\ref{appendix:codec_metrics}.

\textbf{Baselines:} 
We compare our codec against some state-of-the-art (SOTA) baselines, 
including DAC~\citep{dac}, Encodec~\citep{defossez2022high}, X-Codec~\citep{ye2024codec}, 
X-Codec2~\citep{xcodec2}, BiCodec~\citep{SparkTTS}, WavTokenizer~\citep{wavtokenizer}, 
and UniCodec~\citep{unicodec}.
All results are obtained using their official checkpoints.

\subsubsection{Experimental Results}

\begin{table}[ht]
	\centering
	\caption{Comparison between different codec models on LibriSpeech test-clean set, 
	where \textbf{FPS} and \textbf{BPS} denotes frame per second and bitrate per second, respectively. 
	\textbf{Nq} represents the number of codebook layer. 
	\textbf{Unified} indicates whether the model supports general audio or only speech.}
	\label{tab:codec1}
	\resizebox{\columnwidth}{!}{
	\begin{tabular}{lcccc | *{5}{c}}
		\toprule
		\textbf{Model} & \textbf{Unified} & \textbf{FPS} & \textbf{Nq} & \textbf{BPS} & \textbf{PESQ}($\uparrow$) & \textbf{STOI}($\uparrow$) & \textbf{UTMOS}($\uparrow$) & \textbf{SPK-SIM}($\uparrow$) & \textbf{WER}($\downarrow$) \\
		\midrule
		\midrule
		Ground Truth & - & - & - & - & 4.64 & 1.00 & 4.09 & 1.00 & 2.43 \\
		\midrule
		Encodec & \ding{55} & 75 & 8 & 6000 & \textbf{2.77} & \textbf{0.94} & 3.09 & \textbf{0.89} & \textbf{2.64} \\
		X-Codec & \ding{55} & 50 & 4 & 2000 & \textbf{2.77} & 0.87 & \textbf{4.21} & 0.72 & 3.13 \\
		WavTokenizer & \ding{55} & 75 & 1 & 900 & 2.39 & 0.91 & 4.00 & 0.68 & 5.43 \\
		X-Codec2 & \ding{55} & 50 & 1 & 800 & 2.43 & 0.92 & 4.13 & 0.82 & 3.53 \\
		BiCodec & \ding{55} & 50 & 1 & 650 & 2.51 & 0.92 & 4.18 & 0.80 & 3.23 \\
		\midrule
		DAC & \ding{51} & 50 & 4 & 2000 & 1.42 & 0.84 & 1.83 & 0.60 & 4.32 \\
		X-Codec & \ding{51} & 50 & 4 & 2000 & 2.64 & 0.92 & 3.88 & 0.77 & 3.33 \\
		UniCodec & \ding{51} & 75 & 1 & 900 & 2.56 & 0.92 & 4.00 & 0.76 & 4.23 \\
		WavTokenizer & \ding{51} & 40 & 1 & 480 & 1.88 & 0.87 & 3.78 & 0.57 & 10.03 \\
		\midrule
		H-Codec (ours) & \ding{51} & 25+25 & 4 & 2000 & \textbf{2.99} & \textbf{0.94} & \textbf{4.06} & \textbf{0.84} & \textbf{3.18} \\
		\bottomrule
	\end{tabular}
	}
\end{table}

\textbf{Speech Reconstruction Performance:} 
As reported in \Cref{tab:codec1}, our H-Codec achieves competitive performance at a frame rate of 50. 
Since multi-layer tokens can be predicted simultaneously within a single time step in downstream audio LM, 
we argue that frame rate is more critical, as the number of time steps significantly affects computational cost.
Compared to baselines wich support general audio, H-Codec exhibits better signal reconstruction quality 
(PESQ and STOI), speech naturalness (UTMOS), speaker consistency (SPK-SIM), and 
semantic information preservation (WER). Note that some models achieve higher UTMOS than the ground truth, 
this can be attributed to the generative ability of codec decoder, 
which generates plausible speech at the expense of inacurrate signal alignment. 
Our H-Codec reports UTMOS closely matches that of the ground truth, 
indicating the high fidelity of the reconstructed speech.

\begin{table}[ht]
	\centering
	\caption{Comparison between different codec models on speech (LibriSpeech test-clean), 
	music (MUSDB18-HQ test), and audio (AudioSet eval) domain in terms of Mel loss and STFT loss.}
	\label{tab:codec2}
	\resizebox{\columnwidth}{!}{
	\begin{tabular}{l c *{6}{c}}
		\toprule
		\multirow{2}{*}{\textbf{Model}} & \multirow{2}{*}{\textbf{BPS}} & \multicolumn{2}{c}{\textbf{Speech}} & \multicolumn{2}{c}{\textbf{Music}} & \multicolumn{2}{c}{\textbf{Audio}} \\
		\cmidrule(lr){3-4} \cmidrule(lr){5-6} \cmidrule(lr){7-8}
		& & \textbf{Mel loss}($\downarrow$) & \textbf{STFT loss} ($\downarrow$) & \textbf{Mel loss}($\downarrow$) & \textbf{STFT loss} ($\downarrow$) & \textbf{Mel loss}($\downarrow$) & \textbf{STFT loss} ($\downarrow$) \\
		\midrule
		\midrule
		DAC  & 2000 & 0.6436 & 0.1667 & 0.8443	& 0.2308 & 1.9054 & 0.5164 \\
		X-Codec  &2000 & 0.4225	& 0.1161	& 0.6403	& 0.1804	& 1.5073	& 0.4193 \\
		UniCodec & 900 & 0.4147	& 0.1201 & 0.6488	& 0.1999	& 1.5403	& 0.4760 \\
		WavTokenizer & 480 & 0.5143	& 0.1364	& 0.8174	& 0.2270	& 1.8912	& 0.5201 \\
		\midrule
		H-Codec (ours)  & 2000 & \textbf{0.3394} & \textbf{0.1033} & \textbf{0.5158}  & \textbf{0.1667} & \textbf{1.2512} & \textbf{0.4070} \\
		\bottomrule
	\end{tabular}
	}
\end{table}

\textbf{Audio Reconstruction Performance:} 
\Cref{tab:codec2} presents a comprehensive comparison of audio codec models on speech, music, and general audio tasks. 
All baselines supports general audio reconstruction.
Notably, H-Codec achieves lowest Mel loss and STFT loss on all domain, illustrating the powerful multi-domain reconstruction ability.
This ensures the potential of H-Codec for extensive downstream tasks, including speech, music, and audio generation.

\subsection{UniTok-Audio}

\textbf{Training Datasets:} 
For the training of speech tasks, 
we adopt clean speech samples from the VoxBox~\citep{SparkTTS} dataset, including approximately 3.8k hours of data from 
LibriSpeech~\citep{Librispeech}, MLS\_English~\citep{MLS} and Emilia\_ZH~\citep{Emilia} subset. 
The noise corpus comprises approximately 460 hours of data from the DNS Challenge~\citep{DNS}, 
FSD50K~\citep{FSD50K}, WHAM!~\citep{wichern2019wham}, DESED~\citep{DESED}, DEMAND~\citep{DEMAND}, MUSAN~\citep{MUSAN}, 
DISCO~\citep{DISCO}, MUSDB18-HQ~\citep{MUSDB18HQ}, and TUT Urban Acoustic Scenes~\citep{UAS}. 
We include 60k room impulse response (RIR) samples from SLR28~\citep{slr28} to simulate reverberation.
For the audio data, we include captioned audio samples from 
WavCaps~\citep{wavcaps}, CLAP\_FreeSound~\citep{clap}, VGGSound~\citep{chen2020vggsound}, and Internal data, resulting in approximately 40k hours. 
The simulation pipeline of training samples for all operational modes are 
described in Appendix~\ref{appendix:data_simulation}.

\textbf{Implementation Details:} 
There are 16 layers with 16 attention heads and a hidden dimension of 1024 in the LLaMA-based LM backbone, 
resulting in 481M trainable parameters. 
We also explore different model size configurations in Appendix~\ref{appendix:model_size}.
Our model is trained using AdamW optimizer with 30 epochs, 
where the learning rate reaches a peak of 0.001 after 4000 warm-up steps and reduces at a decay factor of 0.98 in each epoch. 
The lengths of reference audio and input signal are set to 5 seconds for both training and inference phases. 
We train the multi-task version (UniTok-Audio\textsubscript{omni}) and single-task version of UniTok-Audio for performance evaluation. 
For the former, one of the five operational modes is randomly selected for every batch during training. 
For the latter, we report results of models trained within single task.
We also attempt to adopt WavLM\footnote{https://huggingface.co/microsoft/wavlm-base-plus} as the audio encoder for the single-task version. 
Subscripts are used to distinguish different models (e.g., HuBERT-based and WavLM-based single-task verisons for SR are denoted as UniTok-Audio\textsubscript{sr-hubert} and UniTok-Audio\textsubscript{sr-wavlm}).

\textbf{Evaluation Metrics:} 
We adopt multiple evaluation metrics to assess different aspects of the generated audio across tasks. 
For speech tasks, we evaluate quality by DNSMOS (SIG, BAK, OVRL) and NISQA, speaker similarity by SIM, 
intelligibility by WER, and continuity by PLCMOS. 
For the LASS task, we utilize FAD, CLAPScore, and CLAPScore$_A$ to measure the audio separation performance. 
Details about evaluation metrics can be found in Appendix~\ref{appendix:task_metrics}.

\subsubsection{SR Performance}

\begin{table}[ht]
	\centering
	\caption{DNSMOS scores on the Interspeech 2020 DNS Challenge blind test set. 
    ``D" represents discriminative approaches. 
    ``G\textsubscript{c}'' and ``G\textsubscript{d}'' denote generative methods in the continuous domain and discrete domain, respectively.    
    ``No Reverb" subset contains only noise while ``With Reverb" subset additionally involves reverberation. 
    }
	\label{tab:Noise Suppression}
	\small
	\setlength\tabcolsep{3pt}
	\begin{tabular}{ll ccc ccc}
		\toprule
		\multirow{2}{*}{\textbf{Model}} & \multirow{2}{*}{\textbf{Type}} & \multicolumn{3}{c}{\textbf{With Reverb}} & \multicolumn{3}{c}{\textbf{No Reverb}} \\
		\cmidrule (lr){3-5} \cmidrule (lr){6-8}
		& & \textbf{SIG}($\uparrow$) & \textbf{BAK}($\uparrow$) & \textbf{OVRL}($\uparrow$) & \textbf{SIG}($\uparrow$) & \textbf{BAK}($\uparrow$) & \textbf{OVRL}($\uparrow$) \\
		\midrule
        \midrule
		Noisy    & -  & 1.76 & 1.50 & 1.39 & 3.39 & 2.62 & 2.48 \\
		\midrule
		Conv-TasNet & D & 2.42 & 2.71 & 2.01 & 3.09 & 3.34 & 3.00 \\
		DEMUCS & D & 2.86 & 3.90 & 2.55 & 3.58 & 4.15 & 3.35 \\
		FRCRN & D & 2.93 & 2.92 & 2.28 & 3.58 & 4.13 & 3.34 \\
		\midrule
        FlowSE & G\textsubscript{c} & 3.60 & 4.10 & 3.33 & 3.69 & 4.20 & 3.45 \\
        UniFlow & G\textsubscript{c} & 3.59 & 4.12 & 3.32 & \textbf{3.72} & \textbf{4.21} & \textbf{3.48} \\
        \midrule
		SELM & G\textsubscript{d} & 3.16 & 3.58 & 2.70 & 3.51 & 4.10 & 3.26 \\
		MaskSR & G\textsubscript{d} & 3.53 & 4.07 & 3.25 & 3.59 & 4.12 & 3.34 \\
		AnyEnhance & G\textsubscript{d} & 3.50 & 4.04 & 3.20 & 3.64 & 4.18 & 3.42 \\
		GenSE & G\textsubscript{d} & 3.49 & 3.73 & 3.19 & 3.65 & 4.18 & 3.43 \\
		Metis-SE & G\textsubscript{d} & \textbf{3.68} & \textbf{4.14} & \textbf{3.44} & 3.64 & 4.17 & 3.43 \\
		LLaSE-G1 & G\textsubscript{d} & 3.59 & 4.10 & 3.33 & 3.66 & 4.17 & 3.42 \\
		UniSE & G\textsubscript{d} & 3.67 & 4.10 & 3.40 & 3.67 & 4.14 & 3.43 \\
		\midrule
		UniTok-Audio\textsubscript{sr-hubert} & G\textsubscript{d} & 3.67 & 4.11 & 3.40 & 3.66 & 4.15 & 3.41 \\
		UniTok-Audio\textsubscript{sr-wavlm} & G\textsubscript{d} & 3.67 & 4.10 & 3.40 & 3.66 & 4.14 & 3.42 \\
		UniTok-Audio\textsubscript{omni} & G\textsubscript{d} & 3.67 & 4.12 & 3.42 & 3.66 & 4.15 & 3.44 \\
		\bottomrule
	\end{tabular}
\end{table}

\begin{table}[ht]
	\centering
	\caption{DNSMOS OVRL and PLCMOS scores on 2022 ICASSP PLC challenge blind test set.}
	\label{tab:PLC scores}
	\small
	\setlength\tabcolsep{5pt}
	\begin{tabular}{ll cc}
		\toprule
		\textbf{Model} & \textbf{Type}  & {\textbf{OVRL}}($\uparrow$) & {\textbf{PLCMOS}}($\uparrow$) \\
		\midrule
        \midrule
		Noisy    & -  & 2.56 & 2.90 \\
		\midrule
		KuaishouNet~\citep{2022kuaishou}   & D  &  -  & 4.27\\
		LPCNet~\citep{LPCNet}   & D  &   3.09   & 3.74\\
		PLCNet~\citep{liu2022plcnet}   & D  &   -    & 3.83\\
		BS-PLCNet~\citep{BS-PLCNet}   & D  & 3.20 & 4.29 \\
		LLaSE-G1~\citep{kang2025llaseg1} & G\textsubscript{d}  & 3.03 & 3.68 \\
		\midrule
		UniTok-Audio\textsubscript{sr-hubert} & G\textsubscript{d}  & 3.30 & 4.55 \\
		UniTok-Audio\textsubscript{sr-wavlm} & G\textsubscript{d}  & 3.33 & 4.55 \\
		UniTok-Audio\textsubscript{omni} & G\textsubscript{d}  & \textbf{3.35} & \textbf{4.58} \\
		\bottomrule
	\end{tabular}
\end{table}

\textbf{Evaluation Configuration:} 
We evaluate speech restoration performance on the synthetic test sets of 2020 DNS Challenge~\citep{DNS} (including ``With Reverb'' and ``No Reverb'') and 
2022 PLC Challenge~\citep{plc_challenge} blind test set. 
Baselines include Conv-TasNet~\citep{ConvTasNet}, DEMUCS~\citep{demucs}, FRCRN~\citep{frcrn}, FlowSE~\citep{lee2025flowse}, 
UniFlow~\citep{wang2025uniflow}, SELM~\citep{wang2024selm}, MaskSR~\citep{li2024masksr}, AnyEnhance~\citep{zhang2025anyenhance}, 
GenSE~\citep{yao2025gense}, Metis-SE~\citep{wang2025metis}, LLaSE-G1~\citep{kang2025llaseg1}, 
UniSE~\citep{unise}, KuaishouNet~\citep{2022kuaishou}, LPCNet~\citep{LPCNet}, PLCNet~\citep{liu2022plcnet}, 
and BS-PLCNet~\citep{BS-PLCNet}. 

\textbf{Results:} 
\Cref{tab:Noise Suppression} presents the SR performance comparison on 2020 DNS Challenge test sets. 
It is clear that generative models generally outperform discriminative ones. 
Continuous-domain generative approaches perform well on the ``No Reverb'' subset, 
highlighting the potential of continuous methods in terms of generated signal quality. 
However, discrete-domain generative approaches can perform better under reverberant conditions, indicating 
that discrete representations may simplify the modeling difficulty of reverberation components. 
Our UniTok-Audio achieves comparable performance among SOTA baselines, 
and the single-task versions with different audio encoders result in similar performance to UniTok-Audio\textsubscript{omni}. 
In addition, \Cref{tab:PLC scores} reports the performance on packet loss concealment (PLC), 
a subtask of SR aimed at recovering speech frames lost during transmission. 
UniTok-Audio surpasses baselines in terms of both signal quality and continuity, 
showing powerful content understanding and generation capabilities of the framework.



\subsubsection{TSE Performance}

\begin{table}[ht]
	\centering
	\caption{TSE results on Libri2Mix clean test set.}
	\label{tab:tse}
	\small
	\setlength{\tabcolsep}{4pt}
	\begin{tabular}{lcccccc}
		\toprule
		\textbf{Model} & \textbf{Type} & \textbf{SIG}($\uparrow$) & \textbf{BAK}($\uparrow$) & \textbf{OVRL}($\uparrow$) & \textbf{NISQA}($\uparrow$) & \textbf{SIM}($\uparrow$) \\
		\midrule
		Mixture & - & 3.38 & 3.10 & 2.65 & 2.45 & 0.85 \\
		\midrule
        \midrule
		Spex+ & D & 3.38 & 3.77 & 3.00 & 3.03 & 0.96 \\
		WeSep & D & 3.56 & 3.93 & 3.23 & 4.04 & \textbf{0.99} \\
		TSELM-L & G\textsubscript{d} & 3.55 & \textbf{4.08} & 3.23 & 4.03 & 0.91 \\
		AnyEnhance & G\textsubscript{d} & 3.64 & 4.07 & \textbf{3.35} & 4.28 & 0.91 \\
		LLaSE-G1 & G\textsubscript{d} & 3.53 & 4.01 & 3.22 & 3.89 & 0.92 \\
		Metis-TSE & G\textsubscript{d} & \textbf{3.65} & \textbf{4.08} & 3.34 & \textbf{4.36} & - \\
		LauraTSE & G\textsubscript{d} & 3.61 & \textbf{4.08} & 3.34 & 4.33 & 0.97 \\
		UniSE & G\textsubscript{d} &3.62 & 4.06 &3.33 &4.00 &0.95 \\
		\midrule
		UniTok-Audio\textsubscript{tse-hubert} & G\textsubscript{d} & 3.58 & 4.03 & 3.31 & 3.97 & 0.95 \\
		UniTok-Audio\textsubscript{tse-wavlm} & G\textsubscript{d} & 3.60 & 4.04 & 3.32 & 3.99 & 0.95 \\
		UniTok-Audio\textsubscript{omni} & G\textsubscript{d} & 3.62 & 4.05 & 3.32 & 4.00 & 0.95 \\
		\bottomrule
	\end{tabular}
\end{table}

\textbf{Evaluation Configuration:} 
The performance of TSE is evaluated on the Libri2Mix~\citep{librimix} clean test set. 
Baselines include Spex+~\citep{spex}, WeSep~\citep{WeSep}, TSELM-L~\citep{TSELM}, 
AnyEnhance~\citep{zhang2025anyenhance}, LLaSE-G1~\citep{kang2025llaseg1}, 
Metis-TSE ~\citep{wang2025metis}, LauraTSE~\citep{LauraTSE}, and UniSE~\citep{unise}.

\textbf{Results:} 
\Cref{tab:tse} shows the performance comparison for TSE task. 
The results indicate that generative methods achieve higher speech quality than discriminative approaches 
but struggle with speaker similarity. 
This can be attributed to the upper bound limitation of codecs' reconstruction fidelity~\citep{unise}. 
Our UniTok-Audio maintains comparable performance compared to SOTA baselines, 
demonstrating the feasibility of constructing a unified framework.

\subsubsection{SS Performance}

\begin{table}[tt]
	\centering
	\caption{SS results on Libri2Mix and WSJ0-2mix test sets.}
    \label{tab:ss}
	\resizebox{\columnwidth}{!}{
		\begin{tabular}{lccccccc}
			\toprule
			\multirow{2}{*}{\textbf{Model}} & \multirow{2}{*}{\textbf{Type}} & \multicolumn{3}{c}{\textbf{Libri2Mix}} & \multicolumn{3}{c}{\textbf{WSJ0-2mix}} \\
			\cmidrule(lr){3-5} \cmidrule(lr){6-8}
			& & \textbf{SIG}($\uparrow$) &	\textbf{BAK}($\uparrow$) &	\textbf{OVRL}($\uparrow$) & \textbf{SIG}($\uparrow$) & \textbf{BAK}($\uparrow$) & \textbf{OVRL}($\uparrow$) \\
			\midrule
			\midrule
			Mixture & - & 2.33 & 1.66 & 1.64 & 3.42 & 3.20 & 2.76 \\
            \midrule
			Sepformer~\citep{Sepformer} & D & 3.33 & 3.88 & 3.02 & 3.43 & 3.96 & 3.14 \\
			Mossformer2~\citep{MossFormer2} & D & 3.44 & 3.94 & 3.11 & 3.50 & \textbf{4.05} & 3.23 \\
			LLaSE-G1~\citep{kang2025llaseg1} & G\textsubscript{d} & 3.48 & 3.83 & 3.11 & 3.52 & 3.92 & 3.19 \\
			\midrule
			UniTok-Audio\textsubscript{omni} & G\textsubscript{d} & \textbf{3.56} & \textbf{4.04} & \textbf{3.25} & \textbf{3.57} & 3.96 & \textbf{3.26} \\
			\bottomrule
		\end{tabular}
	}
	\vspace{-0.2em}
\end{table}

\textbf{Evaluation Configuration:} 
We evaluate SS performance on Libri2Mix noisy test set and WSJ0-2mix~\citep{wsj0mix} test set, 
where the former additionally evaluates the denoising ability of models.  
Baselines include Sepformer~\citep{Sepformer}, Mossformer2~\citep{MossFormer2}, and LLaSE-G1~\citep{kang2025llaseg1}. 

\textbf{Results:} 
\Cref{tab:ss} reports the performance comparison for SS task, showing that our model achieves superior performance than baselines. 
This verifies the effectiveness of our iterative inference strategy in handling the SS task that requires multiple output tracks. 
Note that although the experiments are conducted with the 2-speaker configuration, 
our approach can be extended to scenarios with more sources when the target signal of rTSE mode is defined as all remaining speakers. 
The single-task version is not reported since the inference phase of SS requires the cooperation of multiple modes.



\subsubsection{VC Performance}

\begin{table}
	\caption{Performance comparison on the VC task.}
	\label{tab:results-vc}
	\centering
		\begin{threeparttable}
			\begin{tabular}{lccccc}
				\toprule
				\textbf{Model} & \textbf{Type} & \textbf{WER}($\downarrow$) & \textbf{SIM}($\uparrow$) & \textbf{DNSMOS}($\uparrow$) & \textbf{NISQA}($\uparrow$) \\
				\midrule
				\midrule
				HierSpeech++ & G\textsubscript{c} & 4.87 & 0.38 & 3.40 & 3.79 \\
				LM-VC & G\textsubscript{d} & 8.35 & 0.29 & 3.46 & 3.93 \\
				UniAudio & G\textsubscript{d} & 9.00 & 0.25 & 3.47 & 4.28 \\
				Vevo & G\textsubscript{c} & 3.48 & 0.38 & 3.47 & 4.30 \\
				Metis-VC & G\textsubscript{d} & 4.49 & 0.50 & 3.48 & 4.46 \\
				\midrule
				UniTok-Audio\textsubscript{vc-hubert} & G\textsubscript{d} & 4.15 & 0.48 & 3.42 & 4.43 \\
				UniTok-Audio\textsubscript{vc-wavlm} & G\textsubscript{d} & \textbf{3.02} & \textbf{0.51} & 3.46 & 4.46 \\
				UniTok-Audio\textsubscript{omni} & G\textsubscript{d} & 4.23  & 0.50 & \textbf{3.51} & \textbf{4.51} \\
				\bottomrule
			\end{tabular}
		\end{threeparttable}%
	\vspace{-5mm}
\end{table}

\textbf{Evaluation Configuration:} 
Following~\citep{wang2025metis}, we create test set for the VC task using VCTK~\citep{veaux2017cstr} dataset. 
We randomly select 200 recordings from the dataset as source speech, 
and for each source sample, a sample from another speaker is picked as the reference speech.
Baselines include HierSpeech++~\citep{lee2023hierspeech++}, LM-VC~\citep{wang2023lm}, UniAudio~\citep{yang2024uniaudio}, 
Vevo~\citep{vevo}, and Metis~\citep{wang2025metis}.

\textbf{Results:} 
VC results are presented in \Cref{tab:results-vc}, showing the superiority of UniTok-Audio in 
speech quality, speaker similarity, and intelligibility. 
We observe that UniTok-Audio\textsubscript{vc-wavlm} outperforms UniTok-Audio\textsubscript{vc-hubert}, indicating that 
WavLM performs better in extracting semantic information and speaker characteristics. 
The performance degrades when extending to multiple tasks from single-task version, 
implying the distinct pattern between VC and other tasks, 
where the former changes the property of the input signal rather than restoring or extracting certain components.

\subsubsection{LASS Performance}

\begin{table}[ht]
	\centering
	\caption{LASS results on 2024 DCASE LASS validation set. 
    }
	\label{tab:lass_results}
	\begin{tabular}{lcccc}
		\toprule
		\textbf{Model} & \textbf{Type} & \textbf{FAD}($\downarrow$) & \textbf{CLAPScore}($\uparrow$) & \textbf{CLAPScore}$_A$($\uparrow$)  \\
		\midrule
        \midrule
		Mixture & - & - & 23.83 & 60.39 \\
        \midrule
		LASS-Net & D & 2.57 & 23.04 & 65.14 \\
		FlowSep & G\textsubscript{c} & \textbf{0.50} & 20.00 & 63.47 \\
		\midrule 
		UniTok-Audio\textsubscript{lass-hubert} & G\textsubscript{d} & 0.68 & \textbf{28.85} & \textbf{65.56} \\ 
		UniTok-Audio\textsubscript{omni} & G\textsubscript{d} & 1.48 & 26.21 & 61.21 \\ 
		\bottomrule
	\end{tabular}
\end{table}

\textbf{Evaluation Configuration:} 
We adopt 2024 DCASE LASS\footnote{https://dcase.community/challenge2024/task-language-queried-audio-source-separation} validation set 
to evaluate the LASS performance, which contains 3k synthetic mixtures mixed from 1k audio clips. 
Baselines include LASS-Net~\citep{lass-net} and FlowSep~\citep{yuan2025flowsep}. 

\textbf{Results:} 
As shown in \Cref{tab:lass_results}, UniTok-Audio achieves competitive performance in the LASS task, 
indicating effective exploitation of textual information. 
We prove that the unified domain codec has potential to handle the LASS tasks. 
The single-task version outperforms UniTok-Audio\textsubscript{omni}, 
which can be attributed to the domain gap between speech and audio.

\section{Conclusion}
\label{sec:conclusion}
In this work, we propose UniTok-Audio, 
a framework that resembles multiple time-aligned audio tasks. 
We uniify different learning patterns of multiple tasks in a single framework using a special task token, 
which indicates current operational mode of model. 
This paper also introduces H-Codec, achieving high-fidelity reconstruction quality with dual-stream architecture 
that quantize acoustic and semantic features simultaneously. 
Based on H-Codec, UniTok-Audio adopts continuous conditional embeddings to generates multi-layer discrete tokens in parallel.
Extensive experiments demonstrate that UniTok-Audio achieves competitive performance across diverse tasks 
with limited training data and moderate model size, 
highlighting its potential as a foundation model for unified AR audio generation.
\newpage

\bibliography{iclr2026_conference}
\bibliographystyle{iclr2026_conference}

\appendix

\section{Data Simulation}
\label{appendix:data_simulation}
A data simulation pipeline is designed to synthesis data pairs dynamically during training. 
Considering various types of degradation in the SR task, 
we apply multiple distortions to a speech sample with independent probabilities, 
where the distortion categories and corresponding configurations are shown in ~\Cref{tab:simu}. 
The distortion chain is also applied to the TSE and rTSE modes,
except that the probability of interfering speaker is set to 1.0 and the SIR is uniformly sampled between 
-5 and 5 dB. 
For the LASS mode, we mix the target audio with another randomly selected audio using a SIR ranges from -5 to 20 dB. 
For the VC mode, we leverage a voice conversion model\footnote{https://github.com/myshell-ai/OpenVoice} to perform timbre perturbation using randomly selected target speech and reference speech, 
generating 6k hours of fixed training dataset. 
The perturbed sample is used as input to predict the target speech based on another speech of the target speaker.


\begin{table}[htbp]
	\centering
	\small
	\caption{Distortion categories and corresponding configurations, 
    where SNR and SIR denote the signal-to-noise ratio and signal-to-interference ratio, respectively.
    }
	\label{tab:simu}
	\begin{tabular}{lcc}
		\toprule
		\textbf{Distortion} & \textbf{Occurrence Probability} & \textbf{Hyperparameters} \\
		\midrule
        \midrule
		Additive Noise & 0.5 & SNR $\sim$ Uniform([-15, 20]) dB \\
		\midrule
		Reverberation & 0.4 & - \\
		\midrule
		Clipping & 0.3 & 
		\begin{tabular}{@{}c@{}}
			Min\_quantile $\sim$ Uniform([0.0, 0.1]) \\
			Max\_quantile $\sim$ Uniform([0.9, 1.0])
		\end{tabular} \\
		\midrule
		Bandwidth Limitation & 0.3 & Cutoff frequencies $\in$ \{2, 4\} kHz \\
		\midrule
		Packet Loss & 0.3 & Loss rate $\sim$ Uniform([0.05, 0.25]) \\
		\midrule
		Interfering Speaker & 0.2 & SIR $\sim$ Uniform([15, 25]) dB \\
		\bottomrule
	\end{tabular}
\end{table}

\section{Evaluation Metrics}
\label{appendix:evaluation_metrics}

\subsection{Codec Metrics}
\label{appendix:codec_metrics}
\textbf{PESQ}~\citep{rix2001perceptual}: 
The perceptual evaluation of speech quality (PESQ) assesses perceptual speech quality by comparing 
the reconstructed speech to the ground-truth target speech signal. 
We employ the wideband PESQ scoring from 1 (poor) to 4.5 (excellent).

\textbf{STOI}~\citep{andersen2017non}: 
The short-time objective intelligibility (STOI) evaluates the intelligibility of speech signals, ranging from 0 to 1. 
The higher STOI score indicates a higher intelligibility and better preservation of the speech content.

\textbf{UTMOS}~\citep{saeki2022utmos}: 
An automatic Mean Opinion Score (MOS) predictor\footnote{https://github.com/tarepan/SpeechMOS} measuring the naturalness of speech. 

\textbf{WER}: 
Word Error Rate (WER) measures the intelligibility of the generated speech by using the automatic speech recognition (ASR) model. 
We utilize a HuBERT-based ASR system\footnote{https://huggingface.co/facebook/hubert-large-ls960-ft} to calculate WER. 

\textbf{SPK-SIM}: 
A WavLM-based speaker verification model\footnote{https://github.com/microsoft/UniSpeech/tree/main/downstreams/speaker\_verification} 
is used to calculate the speaker similarity between the reconstructed speech and target speech.

\textbf{STFT Loss \& Mel Loss}: 
We calculate the L1 loss between the magnitude spectrum of the reconstructed speech and target speech, 
where the STFT is performed using a Hann window with a length of 1024 and a shift of 256.
For the Mel Loss, 100 mel filters are utilized.

\subsection{Audio Task Metrics}
\label{appendix:task_metrics}

\textbf{DNSMOS}~\citep{reddy2022dnsmosp835nonintrusiveperceptual}: 
DNSMOS is a neural network-based MOS estimator\footnote{https://github.com/microsoft/DNS-Challenge/tree/master/DNSMOS} that correlates strongly with human quality ratings. 
It comprises three components: 1) speech quality (\textbf{SIG}), 2) background noise quality (\textbf{BAK}), and 3) overall quality (\textbf{OVRL}). 
Note that for the VC task, DNSMOS scores are calculated by averaging three components. 

\textbf{NISQA}~\citep{mittag2021nisqa}: 
NISQA\footnote{https://github.com/gabrielmittag/NISQA} is a deep learning framework for speech quality prediction. 
We report NISQA for the TSE and VC tasks. 

\textbf{SIM}: 
For the TSE task, we evaluate the speaker similarity using finetuned WavLM-base\footnote{https://huggingface.co/microsoft/wavlm-base-plus-sv} following \citep{LauraTSE}. 
While for the VC task, speaker embeddings are computed using the WavLM TDNN\footnote{https://github.com/microsoft/UniSpeech/tree/main/downstreams/speaker\_verification}. 

\textbf{WER}: 
We utilize the whisper-large-v3\footnote{https://huggingface.co/openai/whisper-large-v3}~\citep{radford2023robust} 
to obtain the transcriptions of converted speech in the VC task, thereby calculating WER with 
the ground-truth text of source speech. 

\textbf{PLCMOS}~\citep{DienerPSSAC23}: 
A metric\footnote{https://github.com/microsoft/PLC-Challenge/tree/main/PLCMOS} designed to evaluate the quality of speech enhanced by PLC algorithms, 
outputting a single score ranging from 1 to 5 (higher is better). 

\textbf{FAD}~\citep{kilgour2018fr}: 
Fréchet Audio Distance (FAD)\footnote{https://github.com/gudgud96/frechet-audio-distance} measures the quality of generated audio by comparing the statistics of deep features between real and synthesized audio. 
Lower FAD value indicates higher fidelity and better distributional alignment.

\textbf{CLAPScore \& CLAPScore$_A$}~\citep{clap}: 
CLAPScore measures text-audio similarity using joint embeddings from a contrastive language-audio pretraining (CLAP) model\footnote{https://github.com/LittleFlyingSheep/CLAPScore\_for\_LASS}.
While CLAPScore$_A$ evaluates the similarity between the output audio and the target audio.

\section{Model Size vs. Performance}
\label{appendix:model_size}

\begin{table}[h!]
	\centering
	\caption{Model configurations of different UniTok-Audio versions.}
	\begin{tabular}{lcccc}
		\toprule
		Model Size  & Depth & Embed Size & Num Heads &  Trainable Params \\
		\midrule
        \midrule
		UniTok-Audio-S  &   8    &  768         &8     &109M           \\
		UniTok-Audio &   16    &  1024         &16    &481M           \\
		UniTok-Audio-L  &   44    &  1024        &32   &1.3B       \\
		\bottomrule
	\end{tabular}
	\label{tab:model_config}
\end{table}

\begin{table}
	\caption{VC performance across different model sizes.}
	\centering
		\begin{threeparttable}
			\begin{tabular}{lcccc}
				\toprule
				\textbf{Model} & \textbf{WER}($\downarrow$) & \textbf{SIM}($\uparrow$) & \textbf{DNSMOS}($\uparrow$) & \textbf{NISQA}($\uparrow$) \\
				\midrule
				\midrule
                UniTok-Audio-S & 5.38  & 0.42 & 3.41 & 4.30 \\
				UniTok-Audio & 3.02 & 0.51 & 3.46 & 4.46 \\
				UniTok-Audio-L & \textbf{2.10} & \textbf{0.61} & \textbf{3.61} & \textbf{4.54} \\
				\bottomrule
			\end{tabular}
		\end{threeparttable}%
	\label{tab:model_size_result}
	\vspace{-5mm}
\end{table}

\Cref{tab:model_config} reports the hyperparameter configurations of different UniTok-Audio versions.
UniTok-Audio-S and UniTok-Audio-L denote the small and large version, respectively. 
The VC performance in terms of different verisons are shown in \Cref{tab:model_size_result}, where 
all versions are trained for the single VC task using WavLM-based audio encoder. 
It can be seen that increasing the model size consistently improves performance, 
in accordance with scaling laws. 
This indicates the potential of UniTok-Audio to be extended to a larger model size. 
To balance complexity and performance, we report the medium-sized verison in the main text.

\end{document}

%% file: math_commands.tex
\usepackage{amsmath,amsfonts,bm}

\def\eqref#1{equation~\ref{#1}}

\def\1{\bm{1}}

\DeclareMathAlphabet{\mathsfit}{\encodingdefault}{\sfdefault}{m}{sl}
\SetMathAlphabet{\mathsfit}{bold}{\encodingdefault}{\sfdefault}{bx}{n}













%% file: UniTok-Audio-1029.bbl
\begin{thebibliography}{95}
\providecommand{\natexlab}[1]{#1}
\providecommand{\url}[1]{\texttt{#1}}
\expandafter\ifx\csname urlstyle\endcsname\relax
  \providecommand{\doi}[1]{doi: #1}\else
  \providecommand{\doi}{doi: \begingroup \urlstyle{rm}\Url}\fi

\bibitem[Andersen et~al.(2017)Andersen, de~Haan, Tan, and Jensen]{andersen2017non}
Asger~Heidemann Andersen, Jan~Mark de~Haan, Zheng-Hua Tan, and Jesper Jensen.
\newblock A non-intrusive short-time objective intelligibility measure.
\newblock In \emph{2017 IEEE International Conference on Acoustics, Speech and Signal Processing (ICASSP)}, pp.\  5085--5089. IEEE, 2017.

\bibitem[Chang et~al.(2022)Chang, Zhang, Jiang, Liu, and Freeman]{MaskGIT}
Huiwen Chang, Han Zhang, Lu~Jiang, Ce~Liu, and William~T. Freeman.
\newblock {MaskGIT}: Masked generative image transformer.
\newblock In \emph{Proceedings of the IEEE/CVF conference on computer vision and pattern recognition}, pp.\  11305--11315, 2022.
\newblock \doi{10.1109/CVPR52688.2022.01103}.

\bibitem[Chen et~al.(2020)Chen, Xie, Vedaldi, and Zisserman]{chen2020vggsound}
Honglie Chen, Weidi Xie, Andrea Vedaldi, and Andrew Zisserman.
\newblock {VGGSound}: {A} large-scale audio-visual dataset.
\newblock In \emph{IEEE International Conference on Acoustics, Speech and Signal Processing}, pp.\  721--725. IEEE, 2020.

\bibitem[Chen et~al.(2022)Chen, Wang, Chen, Wu, Liu, Chen, Li, Kanda, Yoshioka, Xiao, et~al.]{chen2022wavlm}
Sanyuan Chen, Chengyi Wang, Zhengyang Chen, Yu~Wu, Shujie Liu, Zhuo Chen, Jinyu Li, Naoyuki Kanda, Takuya Yoshioka, Xiong Xiao, et~al.
\newblock Wavlm: Large-scale self-supervised pre-training for full stack speech processing.
\newblock \emph{IEEE Journal of Selected Topics in Signal Processing}, 16\penalty0 (6):\penalty0 1505--1518, 2022.

\bibitem[Copet et~al.(2023)Copet, Kreuk, Gat, Remez, Kant, Synnaeve, Adi, and D{\'e}fossez]{musicgen}
Jade Copet, Felix Kreuk, Itai Gat, Tal Remez, David Kant, Gabriel Synnaeve, Yossi Adi, and Alexandre D{\'e}fossez.
\newblock Simple and controllable music generation.
\newblock \emph{arXiv preprint arXiv:2306.05284}, 2023.

\bibitem[Cosentino et~al.(2020)Cosentino, Pariente, Cornell, Deleforge, and Vincent]{librimix}
Joris Cosentino, Manuel Pariente, Samuele Cornell, Antoine Deleforge, and Emmanuel Vincent.
\newblock {LibriMix}: An open-source dataset for generalizable speech separation.
\newblock \emph{arXiv preprint arXiv:2005.11262}, 2020.

\bibitem[Defferrard et~al.(2017)Defferrard, Benzi, Vandergheynst, and Bresson]{Defferrard2017FMA}
Micha{\"e}l Defferrard, Kirell Benzi, Pierre Vandergheynst, and Xavier Bresson.
\newblock {FMA}: {A} dataset for music analysis.
\newblock \emph{arXiv preprint arXiv:1612.01840}, 2017.

\bibitem[D{\'e}fossez et~al.(2022{\natexlab{a}})D{\'e}fossez, Copet, Synnaeve, and Adi]{defossez2022high}
Alexandre D{\'e}fossez, Jade Copet, Gabriel Synnaeve, and Yossi Adi.
\newblock High fidelity neural audio compression.
\newblock \emph{arXiv preprint arXiv:2210.13438}, 2022{\natexlab{a}}.

\bibitem[D{\'e}fossez et~al.(2022{\natexlab{b}})D{\'e}fossez, Copet, Synnaeve, and Adi]{encodec}
Alexandre D{\'e}fossez, Jade Copet, Gabriel Synnaeve, and Yossi Adi.
\newblock High fidelity neural audio compression.
\newblock \emph{arXiv preprint arXiv:2210.13438}, 2022{\natexlab{b}}.

\bibitem[D{\'e}fossez et~al.(2024)D{\'e}fossez, Mazar{\'e}, Orsini, Royer, P{\'e}rez, J{\'e}gou, Grave, and Zeghidour]{moshi}
Alexandre D{\'e}fossez, Laurent Mazar{\'e}, Manu Orsini, Am{\'e}lie Royer, Patrick P{\'e}rez, Herv{\'e} J{\'e}gou, Edouard Grave, and Neil Zeghidour.
\newblock Moshi: a speech-text foundation model for real-time dialogue.
\newblock \emph{arXiv preprint arXiv:2410.00037}, 2024.

\bibitem[Diener et~al.(2022)Diener, Sootla, Branets, Saabas, Aichner, and Cutler]{plc_challenge}
Lorenz Diener, Sten Sootla, Solomiya Branets, Ando Saabas, Robert Aichner, and Ross Cutler.
\newblock Interspeech 2022 audio deep packet loss concealment challenge.
\newblock In \emph{Interspeech}, pp.\  580--584, 2022.
\newblock \doi{10.21437/Interspeech.2022-10829}.

\bibitem[Diener et~al.(2023)Diener, Purin, Sootla, Saabas, Aichner, and Cutler]{DienerPSSAC23}
Lorenz Diener, Marju Purin, Sten Sootla, Ando Saabas, Robert Aichner, and Ross Cutler.
\newblock {PLCMOS} - {A} data-driven non-intrusive metric for the evaluation of packet loss concealment algorithms.
\newblock In \emph{{INTERSPEECH}}, pp.\  2533--2537. {ISCA}, 2023.

\bibitem[Défossez et~al.(2019)Défossez, Usunier, Bottou, and Bach]{demucs}
Alexandre Défossez, Nicolas Usunier, Léon Bottou, and Francis Bach.
\newblock Demucs: Deep extractor for music sources with extra unlabeled data remixed.
\newblock \emph{arXiv preprint arXiv:1909.01174}, 2019.

\bibitem[Fonseca et~al.(2022)Fonseca, Favory, Pons, Font, and Serra]{FSD50K}
Eduardo Fonseca, Xavier Favory, Jordi Pons, Frederic Font, and Xavier Serra.
\newblock {FSD50K}: An open dataset of human-labeled sound events.
\newblock \emph{IEEE/ACM Trans. Audio, Speech, Lang. Process.}, 30:\penalty0 829--852, 2022.

\bibitem[Furnon et~al.(2021)Furnon, Serizel, Essid, and Illina]{DISCO}
Nicolas Furnon, Romain Serizel, Slim Essid, and Irina Illina.
\newblock {DNN-based} mask estimation for distributed speech enhancement in spatially unconstrained microphone arrays.
\newblock \emph{IEEE/ACM Trans. Audio, Speech, Lang. Process.}, 29:\penalty0 2310--2323, 2021.

\bibitem[Ge et~al.(2020)Ge, Xu, Wang, Chng, Dang, and Li]{spex}
Meng Ge, Chenglin Xu, Longbiao Wang, Eng~Siong Chng, Jianwu Dang, and Haizhou Li.
\newblock {SpEx+}: A complete time domain speaker extraction network.
\newblock In \emph{Proc. Interspeech}, pp.\  1406--1410, 2020.

\bibitem[Gemmeke et~al.(2017)Gemmeke, Ellis, Freedman, Jansen, Lawrence, Moore, Plakal, and Ritter]{gemmeke2017audio}
Jort~F Gemmeke, Daniel~PW Ellis, Dylan Freedman, Aren Jansen, Wade Lawrence, R~Channing Moore, Manoj Plakal, and Marvin Ritter.
\newblock {AudioSet}: {A}n ontology and human-labeled dataset for audio events.
\newblock In \emph{Proceedings of the Conference of the International Speech Communication Association}, pp.\  776--780. IEEE, 2017.

\bibitem[He et~al.(2024)He, Shang, Wang, Li, Gu, Hua, Liu, Yang, Li, Shi, Wang, Chen, Zhang, and Wu]{Emilia}
Haorui He, Zengqiang Shang, Chaoren Wang, Xuyuan Li, Yicheng Gu, Hua Hua, Liwei Liu, Chen Yang, Jiaqi Li, Peiyang Shi, Yuancheng Wang, Kai Chen, Pengyuan Zhang, and Zhizheng Wu.
\newblock Emilia: An extensive, multilingual, and diverse speech dataset for large-scale speech generation.
\newblock In \emph{Proc. SLT}, pp.\  885--890, 2024.

\bibitem[Hershey et~al.(2016)Hershey, Chen, Le~Roux, and Watanabe]{wsj0mix}
John~R. Hershey, Zhuo Chen, Jonathan Le~Roux, and Shinji Watanabe.
\newblock Deep clustering: Discriminative embeddings for segmentation and separation.
\newblock In \emph{IEEE International Conference on Acoustics, Speech and Signal Processing}, pp.\  31--35, 2016.
\newblock \doi{10.1109/ICASSP.2016.7471631}.

\bibitem[Ho et~al.(2020)Ho, Jain, and Abbeel]{ho2020denoising}
Jonathan Ho, Ajay Jain, and Pieter Abbeel.
\newblock Denoising diffusion probabilistic models.
\newblock \emph{Advances in Neural Information Processing Systems}, 33:\penalty0 6840--6851, 2020.

\bibitem[Hsu et~al.(2021)Hsu, Bolte, Tsai, Lakhotia, Salakhutdinov, and Mohamed]{hubert}
Wei-Ning Hsu, Benjamin Bolte, Yao-Hung~Hubert Tsai, Kushal Lakhotia, Ruslan Salakhutdinov, and Abdelrahman Mohamed.
\newblock Hubert: Self-supervised speech representation learning by masked prediction of hidden units.
\newblock \emph{IEEE/ACM Transactions on Audio, Speech, and Language Processing}, 29:\penalty0 3451--3460, 2021.

\bibitem[Jang et~al.(2021)Jang, Lim, Yoon, Kim, and Kim]{univnet}
Won Jang, Dan Lim, Jaesam Yoon, Bongwan Kim, and Juntae Kim.
\newblock {UnivNet}: A neural vocoder with multi-resolution spectrogram discriminators for high-fidelity waveform generation.
\newblock In \emph{Proceedings of the Conference of the International Speech Communication Association}, pp.\  2207--2211, 2021.
\newblock \doi{10.21437/Interspeech.2021-1016}.

\bibitem[Ji et~al.(2024)Ji, Jiang, Wang, Chen, Fang, Zuo, Yang, Cheng, Wang, Li, et~al.]{wavtokenizer}
Shengpeng Ji, Ziyue Jiang, Wen Wang, Yifu Chen, Minghui Fang, Jialong Zuo, Qian Yang, Xize Cheng, Zehan Wang, Ruiqi Li, et~al.
\newblock Wavtokenizer: an efficient acoustic discrete codec tokenizer for audio language modeling.
\newblock \emph{arXiv preprint arXiv:2408.16532}, 2024.

\bibitem[Jiang et~al.(2025)Jiang, Chen, Ji, Xi, Wang, Zhang, Yue, Zhang, and Li]{unicodec}
Yidi Jiang, Qian Chen, Shengpeng Ji, Yu~Xi, Wen Wang, Chong Zhang, Xianghu Yue, ShiLiang Zhang, and Haizhou Li.
\newblock {U}ni{C}odec: Unified audio codec with single domain-adaptive codebook.
\newblock In Wanxiang Che, Joyce Nabende, Ekaterina Shutova, and Mohammad~Taher Pilehvar (eds.), \emph{Proceedings of the 63rd Annual Meeting of the Association for Computational Linguistics (Volume 1: Long Papers)}, pp.\  19112--19124, Vienna, Austria, July 2025. Association for Computational Linguistics.
\newblock \doi{10.18653/v1/2025.acl-long.937}.
\newblock URL \url{https://aclanthology.org/2025.acl-long.937/}.

\bibitem[Kang et~al.(2025)Kang, Zhu, Zhang, Ye, Liu, Wang, Zhu, Ma, Chen, Xiao, Weng, Xue, and Xie]{kang2025llaseg1}
Boyi Kang, Xinfa Zhu, Zihan Zhang, Zhen Ye, Mingshuai Liu, Ziqian Wang, Yike Zhu, Guobin Ma, Jun Chen, Longshuai Xiao, Chao Weng, Wei Xue, and Lei Xie.
\newblock {LL}a{SE}-g1: Incentivizing generalization capability for {LL}a{MA}-based speech enhancement.
\newblock In \emph{Proceedings of the 63rd Annual Meeting of the Association for Computational Linguistics (Volume 1: Long Papers)}, pp.\  13292--13305, Vienna, Austria, July 2025. Association for Computational Linguistics.
\newblock ISBN 979-8-89176-251-0.
\newblock \doi{10.18653/v1/2025.acl-long.651}.
\newblock URL \url{https://aclanthology.org/2025.acl-long.651/}.

\bibitem[Kilgour et~al.(2018)Kilgour, Zuluaga, Roblek, and Sharifi]{kilgour2018fr}
Kevin Kilgour, Mauricio Zuluaga, Dominik Roblek, and Matthew Sharifi.
\newblock Fr$\backslash$'echet audio distance: A metric for evaluating music enhancement algorithms.
\newblock \emph{arXiv preprint arXiv:1812.08466}, 2018.

\bibitem[Ko et~al.(2017)Ko, Peddinti, Povey, Seltzer, and Khudanpur]{slr28}
Tom Ko, Vijayaditya Peddinti, Daniel Povey, Michael~L. Seltzer, and Sanjeev Khudanpur.
\newblock A study on data augmentation of reverberant speech for robust speech recognition.
\newblock In \emph{IEEE International Conference on Acoustics, Speech and Signal Processing}, pp.\  5220--5224, 2017.
\newblock \doi{10.1109/ICASSP.2017.7953152}.

\bibitem[Kong et~al.(2020)Kong, Kim, and Bae]{hifigan}
Jungil Kong, Jaehyeon Kim, and Jaekyoung Bae.
\newblock {HiFi-GAN}: Generative adversarial networks for efficient and high fidelity speech synthesis.
\newblock In \emph{Advances in Neural Information Processing Systems}, volume~33, pp.\  17022--17033. Curran Associates, Inc., 2020.

\bibitem[Kumar et~al.(2023)Kumar, Seetharaman, Luebs, Kumar, and Kumar]{dac}
Rithesh Kumar, Prem Seetharaman, Alejandro Luebs, Ishaan Kumar, and Kundan Kumar.
\newblock High-fidelity audio compression with improved {RVQGAN}.
\newblock In \emph{Thirty-seventh Conference on Neural Information Processing Systems}, 2023.
\newblock URL \url{https://openreview.net/forum?id=qjnl1QUnFA}.

\bibitem[Le et~al.(2023)Le, Vyas, Shi, Karrer, Sari, Moritz, Williamson, Manohar, Adi, Mahadeokar, and Hsu]{voicebox}
Matthew Le, Apoorv Vyas, Bowen Shi, Brian Karrer, Leda Sari, Rashel Moritz, Mary Williamson, Vimal Manohar, Yossi Adi, Jay Mahadeokar, and Wei-Ning Hsu.
\newblock {Voicebox}: text-guided multilingual universal speech generation at scale.
\newblock In \emph{Proceedings of the 37th International Conference on Neural Information Processing Systems}, NIPS '23, Red Hook, NY, USA, 2023. Curran Associates Inc.

\bibitem[Lee et~al.(2023)Lee, Choi, Kim, and Lee]{lee2023hierspeech++}
Sang-Hoon Lee, Ha-Yeong Choi, Seung-Bin Kim, and Seong-Whan Lee.
\newblock Hierspeech++: Bridging the gap between semantic and acoustic representation of speech by hierarchical variational inference for zero-shot speech synthesis.
\newblock \emph{arXiv preprint arXiv:2311.12454}, 2023.

\bibitem[Lee et~al.(2025)Lee, Cheong, Han, and Shin]{lee2025flowse}
Seonggyu Lee, Sein Cheong, Sangwook Han, and Jong~Won Shin.
\newblock {FlowSE}: {F}low {M}atching-based {S}peech {E}nhancement.
\newblock In \emph{IEEE International Conference on Acoustics, Speech and Signal Processing}, 2025.

\bibitem[Li et~al.(2022)Li, Zheng, Zhang, Guo, and Yu]{2022kuaishou}
Nan Li, Xiguang Zheng, Chen Zhang, Liang Guo, and Bing Yu.
\newblock End-to-end multi-loss training for low delay packet loss concealment.
\newblock In \emph{Interspeech 2022}, pp.\  585--589, 2022.
\newblock \doi{10.21437/Interspeech.2022-11439}.

\bibitem[Li et~al.(2024)Li, Wang, and Liu]{li2024masksr}
Xu~Li, Qirui Wang, and Xiaoyu Liu.
\newblock {MaskSR: Masked Language Model for Full-band Speech Restoration}.
\newblock In \emph{{Interspeech 2024}}, pp.\  2275--2279, 2024.
\newblock \doi{10.21437/Interspeech.2024-1584}.

\bibitem[Lipman et~al.(2023)Lipman, Chen, Ben-Hamu, Nickel, and Le]{lipman2023flow}
Yaron Lipman, Ricky~TQ Chen, Heli Ben-Hamu, Maximilian Nickel, and Matthew Le.
\newblock Flow matching for generative modeling.
\newblock In \emph{Proceedings of the International Conference on Learning Representations}, 2023.

\bibitem[Liu et~al.(2022{\natexlab{a}})Liu, Song, Yang, Yuan, and Wang]{liu2022plcnet}
B.~Liu, Q.~Song, M.~Yang, W.~Yuan, and T.~Wang.
\newblock Plcnet: Realtime packet loss concealment with semi-supervised generative adversarial network.
\newblock In \emph{Interspeech}, pp.\  575--579, 2022{\natexlab{a}}.

\bibitem[Liu et~al.(2024)Liu, Xu, Yuan, Wu, Wang, and Plumbley]{semanticodec}
Haohe Liu, Xuenan Xu, Yi~Yuan, Mengyue Wu, Wenwu Wang, and Mark~D Plumbley.
\newblock Semanticodec: An ultra low bitrate semantic audio codec for general sound.
\newblock \emph{arXiv preprint arXiv:2405.00233}, 2024.

\bibitem[Liu et~al.(2022{\natexlab{b}})]{lass-net}
Xubo Liu et~al.
\newblock Separate what you describe: Language-queried audio source separation.
\newblock In \emph{ICASSP}, pp.\  ... IEEE, 2022{\natexlab{b}}.

\bibitem[Luo \& Mesgarani(2019)Luo and Mesgarani]{ConvTasNet}
Yi~Luo and Nima Mesgarani.
\newblock {Conv-TasNet}: Surpassing ideal time-frequency magnitude masking for speech separation.
\newblock \emph{IEEE/ACM Trans. Audio, Speech, Lang. Process.}, 27\penalty0 (8):\penalty0 1256--1266, 2019.
\newblock \doi{10.1109/TASLP.2019.2915167}.

\bibitem[Mei et~al.(2024)Mei, Meng, Liu, Kong, Ko, Zhao, Plumbley, Zou, and Wang]{wavcaps}
Xinhao Mei, Chutong Meng, Haohe Liu, Qiuqiang Kong, Tom Ko, Chengqi Zhao, Mark~D Plumbley, Yuexian Zou, and Wenwu Wang.
\newblock {WavCaps}: {A} {ChatGPT}-assisted weakly-labelled audio captioning dataset for audio-language multimodal research.
\newblock \emph{IEEE/ACM Transactions on Audio, Speech, and Language Processing}, 32:\penalty0 3339--3354, 2024.

\bibitem[Mentzer et~al.(2024)Mentzer, Minnen, Agustsson, and Tschannen]{fsq}
Fabian Mentzer, David Minnen, Eirikur Agustsson, and Michael Tschannen.
\newblock Finite scalar quantization: {VQ-VAE} made simple.
\newblock In B.~Kim, Y.~Yue, S.~Chaudhuri, K.~Fragkiadaki, M.~Khan, and Y.~Sun (eds.), \emph{International Conference on Representation Learning}, volume 2024, pp.\  51772--51783, 2024.
\newblock URL \url{https://proceedings.iclr.cc/paper_files/paper/2024/file/e2dd53601de57c773343a7cdf09fae1c-Paper-Conference.pdf}.

\bibitem[Mesaros et~al.(2018)Mesaros, Heittola, and Virtanen]{UAS}
Annamaria Mesaros, Toni Heittola, and Tuomas Virtanen.
\newblock A multi-device dataset for urban acoustic scene classification.
\newblock In \emph{Proc. DCASE}, pp.\  9--13, 2018.

\bibitem[Mittag et~al.(2021)Mittag, Naderi, Chehadi, and M{\"o}ller]{mittag2021nisqa}
Gabriel Mittag, Babak Naderi, Assmaa Chehadi, and Sebastian M{\"o}ller.
\newblock Nisqa: A deep cnn-self-attention model for multidimensional speech quality prediction with crowdsourced datasets.
\newblock \emph{arXiv preprint arXiv:2104.09494}, 2021.

\bibitem[Neekhara et~al.(2024)Neekhara, Hussain, Ghosh, Li, and Ginsburg]{t5tts}
Paarth Neekhara, Shehzeen Hussain, Subhankar Ghosh, Jason Li, and Boris Ginsburg.
\newblock Improving robustness of llm-based speech synthesis by learning monotonic alignment.
\newblock In \emph{Interspeech}, pp.\  3425--3429, 2024.
\newblock \doi{10.21437/Interspeech.2024-335}.

\bibitem[Panayotov et~al.(2015)Panayotov, Chen, Povey, and Khudanpur]{Librispeech}
Vassil Panayotov, Guoguo Chen, Daniel Povey, and Sanjeev Khudanpur.
\newblock Librispeech: An asr corpus based on public domain audio books.
\newblock In \emph{Proc. ICASSP}, pp.\  5206--5210, 2015.
\newblock URL \url{https://ieeexplore.ieee.org/document/7178964}.

\bibitem[Parker et~al.(2024)Parker, Smirnov, Pons, Carr, Zukowski, Evans, and Liu]{perceptual_loss}
Julian~D Parker, Anton Smirnov, Jordi Pons, CJ~Carr, Zack Zukowski, Zach Evans, and Xubo Liu.
\newblock Scaling transformers for low-bitrate high-quality speech coding.
\newblock \emph{arXiv preprint arXiv:2411.19842}, 2024.

\bibitem[Peebles \& Xie(2023)Peebles and Xie]{dit}
William Peebles and Saining Xie.
\newblock Scalable diffusion models with {Transformers}.
\newblock In \emph{2023 IEEE/CVF International Conference on Computer Vision (ICCV)}, pp.\  4172--4182, 2023.
\newblock \doi{10.1109/ICCV51070.2023.00387}.

\bibitem[Polyak et~al.(2024)Polyak, Zohar, Brown, Tjandra, Sinha, Lee, Vyas, Shi, Ma, Chuang, et~al.]{polyak2024movie}
Adam Polyak, Amit Zohar, Andrew Brown, Andros Tjandra, Animesh Sinha, Ann Lee, Apoorv Vyas, Bowen Shi, Chih-Yao Ma, Ching-Yao Chuang, et~al.
\newblock Movie {G}en: {A} cast of media foundation models.
\newblock \emph{arXiv preprint arXiv:2410.13720}, 2024.

\bibitem[Pratap et~al.(2020)Pratap, Xu, Sriram, Synnaeve, and Collobert]{MLS}
Vineel Pratap, Qiantong Xu, Anuroop Sriram, Gabriel Synnaeve, and Ronan Collobert.
\newblock {MLS}: A large-scale multilingual dataset for speech research.
\newblock In \emph{Proc. Interspeech}, pp.\  2757--2761, 2020.

\bibitem[Radford et~al.(2023)Radford, Kim, Xu, Brockman, McLeavey, and Sutskever]{radford2023robust}
Alec Radford, Jong~Wook Kim, Tao Xu, Greg Brockman, Christine McLeavey, and Ilya Sutskever.
\newblock Robust speech recognition via large-scale weak supervision.
\newblock In \emph{Proceedings of the International Conference on Machine Learning}, pp.\  28492--28518. PMLR, 2023.

\bibitem[Raffel et~al.(2020)Raffel, Shazeer, Roberts, Lee, Narang, Matena, Zhou, Li, and Liu]{t5}
Colin Raffel, Noam Shazeer, Adam Roberts, Katherine Lee, Sharan Narang, Michael Matena, Yanqi Zhou, Wei Li, and Peter~J. Liu.
\newblock Exploring the limits of transfer learning with a unified text-to-text transformer.
\newblock \emph{Journal of Machine Learning Research}, 21\penalty0 (140):\penalty0 1--67, 2020.
\newblock URL \url{http://jmlr.org/papers/v21/20-074.html}.

\bibitem[Rafii et~al.()Rafii, Liutkus, St{\"o}ter, Mimilakis, and Bittner]{MUSDB18HQ}
Zafar Rafii, Antoine Liutkus, Fabian-Robert St{\"o}ter, Stylianos~Ioannis Mimilakis, and Rachel Bittner.
\newblock {MUSDB18-HQ - an uncompressed version of MUSDB18}.
\newblock [Online].
\newblock Available: \url{https://doi.org/10.5281/zenodo.3338373}.

\bibitem[Reddy et~al.(2020)Reddy, Gopal, Cutler, Beyrami, Cheng, Dubey, Matusevych, Aichner, Aazami, Braun, Rana, Srinivasan, and Gehrke]{DNS}
Chandan K.~A. Reddy, Vishak Gopal, Ross Cutler, Ebrahim Beyrami, Roger Cheng, Harishchandra Dubey, Sergiy Matusevych, Robert Aichner, Ashkan Aazami, Sebastian Braun, Puneet Rana, Sriram Srinivasan, and Johannes Gehrke.
\newblock The interspeech 2020 deep noise suppression challenge: Datasets, subjective testing framework, and challenge results.
\newblock In \emph{Proc. Interspeech}, pp.\  2492--2496, 2020.

\bibitem[Reddy et~al.(2022)Reddy, Gopal, and Cutler]{reddy2022dnsmosp835nonintrusiveperceptual}
Chandan K~A Reddy, Vishak Gopal, and Ross Cutler.
\newblock Dnsmos p.835: A non-intrusive perceptual objective speech quality metric to evaluate noise suppressors, 2022.
\newblock URL \url{https://arxiv.org/abs/2110.01763}.

\bibitem[Rix et~al.(2001)Rix, Beerends, Hollier, and Hekstra]{rix2001perceptual}
Antony~W Rix, John~G Beerends, Michael~P Hollier, and Andries~P Hekstra.
\newblock Perceptual evaluation of speech quality (pesq)-a new method for speech quality assessment of telephone networks and codecs.
\newblock In \emph{IEEE International Conference on Acoustics, Speech and Signal Processing}, volume~2, pp.\  749--752. IEEE, 2001.

\bibitem[Saeki et~al.(2022)Saeki, Xin, Nakata, Koriyama, Takamichi, and Saruwatari]{saeki2022utmos}
Takaaki Saeki, Detai Xin, Wataru Nakata, Tomoki Koriyama, Shinnosuke Takamichi, and Hiroshi Saruwatari.
\newblock Utmos: Utokyo-sarulab system for voicemos challenge 2022.
\newblock \emph{arXiv preprint arXiv:2204.02152}, 2022.

\bibitem[Siuzdak(2024)]{vocos}
Hubert Siuzdak.
\newblock Vocos: Closing the gap between time-domain and fourier-based neural vocoders for high-quality audio synthesis.
\newblock In B.~Kim, Y.~Yue, S.~Chaudhuri, K.~Fragkiadaki, M.~Khan, and Y.~Sun (eds.), \emph{Proceedings of the International Conference on Learning Representations}, volume 2024, pp.\  25719--25733, 2024.
\newblock URL \url{https://proceedings.iclr.cc/paper_files/paper/2024/file/6db0903efdfe9b1bbafb015c10990b78-Paper-Conference.pdf}.

\bibitem[Snyder et~al.(2015)Snyder, Chen, and Povey]{MUSAN}
David Snyder, Guoguo Chen, and Daniel Povey.
\newblock {MUSAN}: A music, speech, and noise corpus.
\newblock \emph{arXiv preprint arXiv:1510.08484}, 2015.

\bibitem[Subakan et~al.(2021)Subakan, Ravanelli, Cornell, Bronzi, and Zhong]{Sepformer}
Cem Subakan, Mirco Ravanelli, Samuele Cornell, Mirko Bronzi, and Jianyuan Zhong.
\newblock Attention is all you need in speech separation.
\newblock In \emph{Proc. ICASSP}, pp.\  21--25, 2021.

\bibitem[Tang et~al.(2024)Tang, Zeng, and Li]{TSELM}
Beilong Tang, Bang Zeng, and Ming Li.
\newblock {TSELM}: Target speaker extraction using discrete tokens and language models.
\newblock \emph{arXiv preprint arXiv:2409.07841}, 2024.

\bibitem[Tang et~al.(2025)Tang, Zeng, and Li]{LauraTSE}
Beilong Tang, Bang Zeng, and Ming Li.
\newblock {LauraTSE}: Target speaker extraction using auto-regressive decoder-only language models.
\newblock \emph{arXiv preprint arXiv:2504.07402}, 2025.

\bibitem[Team(2025)]{qwen3omni}
Qwen Team.
\newblock {Qwen3-Omni} technical report.
\newblock \emph{arXiv preprint arXiv:2509.17765}, 2025.

\bibitem[Thiemann et~al.(2013)Thiemann, Ito, and Vincent]{DEMAND}
Joachim Thiemann, Nobutaka Ito, and Emmanuel Vincent.
\newblock {The diverse environments multi-channel acoustic noise database: A database of multichannel environmental noise recordings}.
\newblock \emph{J. Acoust. Soc. Am.}, 133:\penalty0 3591--3591, 2013.

\bibitem[Touvron et~al.(2023)Touvron, Lavril, Izacard, Martinet, Lachaux, Lacroix, Rozière, Goyal, Hambro, Azhar, Rodriguez, Joulin, Grave, and Lample]{LLaMA}
Hugo Touvron, Thibaut Lavril, Gautier Izacard, Xavier Martinet, Marie-Anne Lachaux, Timothée Lacroix, Baptiste Rozière, Naman Goyal, Eric Hambro, Faisal Azhar, Aurelien Rodriguez, Armand Joulin, Edouard Grave, and Guillaume Lample.
\newblock {LLaMA}: Open and efficient foundation language models.
\newblock \emph{arXiv preprint arXiv:2302.13971}, 2023.

\bibitem[Turpault et~al.(2019)Turpault, Serizel, Salamon, and Shah]{DESED}
Nicolas Turpault, Romain Serizel, Justin Salamon, and Ankit~Parag Shah.
\newblock Sound event detection in domestic environments with weakly labeled data and soundscape synthesis.
\newblock In Michael~I. Mandel, Justin Salamon, and Daniel P.~W. Ellis (eds.), \emph{Proc. DCASE}, pp.\  253--257, 2019.

\bibitem[Valin et~al.(2022)Valin, Mustafa, Montgomery, Terriberry, Klingbeil, Smaragdis, and Krishnaswamy]{LPCNet}
Jean-Marc Valin, Ahmed Mustafa, Christopher Montgomery, Timothy~B. Terriberry, Michael Klingbeil, Paris Smaragdis, and Arvindh Krishnaswamy.
\newblock Real-time packet loss concealment with mixed generative and predictive model, 2022.
\newblock URL \url{https://arxiv.org/abs/2205.05785}.

\bibitem[Vaswani et~al.(2017)Vaswani, Shazeer, Parmar, Uszkoreit, Jones, Gomez, Kaiser, and Polosukhin]{vaswani2017attention}
Ashish Vaswani, Noam Shazeer, Niki Parmar, Jakob Uszkoreit, Llion Jones, Aidan~N Gomez, {\L}ukasz Kaiser, and Illia Polosukhin.
\newblock Attention is all you need.
\newblock \emph{Advances in Neural Information Processing Systems}, 30, 2017.

\bibitem[Veaux et~al.(2017)Veaux, Yamagishi, MacDonald, et~al.]{veaux2017cstr}
Christophe Veaux, Junichi Yamagishi, Kirsten MacDonald, et~al.
\newblock {CSTR VCTK} corpus: {English} multi-speaker corpus for {CSTR} voice cloning toolkit.
\newblock \emph{University of Edinburgh. The Centre for Speech Technology Research (CSTR)}, 6:\penalty0 15, 2017.

\bibitem[Vyas et~al.(2023)Vyas, Shi, Le, Tjandra, Wu, Guo, Zhang, Zhang, Adkins, Ngan, Wang, Cruz, Akula, Akinyemi, Ellis, Moritz, Yungster, Rakotoarison, Tan, Summers, Wood, Lane, Williamson, and Hsu]{audiobox}
Apoorv Vyas, Bowen Shi, Matthew Le, Andros Tjandra, Yi-Chiao Wu, Baishan Guo, Jiemin Zhang, Xinyue Zhang, Robert Adkins, William Ngan, Jeff Wang, Ivan Cruz, Bapi Akula, Akinniyi Akinyemi, Brian Ellis, Rashel Moritz, Yael Yungster, Alice Rakotoarison, Liang Tan, Chris Summers, Carleigh Wood, Joshua Lane, Mary Williamson, and Wei-Ning Hsu.
\newblock Audiobox: Unified audio generation with natural language prompts.
\newblock \emph{arXiv preprint arXiv:2312.15821}, 2023.

\bibitem[Wang et~al.(2023{\natexlab{a}})Wang, Chen, Wu, Zhang, Zhou, Liu, Chen, Liu, Wang, Li, et~al.]{valle}
Chengyi Wang, Sanyuan Chen, Yu~Wu, Ziqiang Zhang, Long Zhou, Shujie Liu, Zhuo Chen, Yanqing Liu, Huaming Wang, Jinyu Li, et~al.
\newblock Neural codec language models are zero-shot text to speech synthesizers.
\newblock \emph{arXiv preprint arXiv:2301.02111}, 2023{\natexlab{a}}.

\bibitem[Wang et~al.(2020)Wang, Tan, and Wang]{over_suppression}
Peidong Wang, Ke~Tan, and De~Liang Wang.
\newblock Bridging the gap between monaural speech enhancement and recognition with distortion-independent acoustic modeling.
\newblock \emph{IEEE/ACM Transactions on Audio, Speech, and Language Processing}, 28:\penalty0 39--48, 2020.
\newblock \doi{10.1109/TASLP.2019.2946789}.

\bibitem[Wang et~al.(2024{\natexlab{a}})Wang, Zhang, Lin, Li, Wang, Ge, Yu, Qian, and Li]{WeSep}
Shuai Wang, Ke~Zhang, Shaoxiong Lin, Junjie Li, Xuefei Wang, Meng Ge, Jianwei Yu, Yanmin Qian, and Haizhou Li.
\newblock {WeSep}: A scalable and flexible toolkit towards generalizable target speaker extraction.
\newblock In \emph{Proc. Interspeech}, pp.\  4273--4277, 2024{\natexlab{a}}.

\bibitem[Wang et~al.(2025{\natexlab{a}})Wang, Jiang, Ma, Zhang, Liu, Li, Liang, Zheng, Wang, Feng, Bian, Ye, Cheng, Yuan, Zhao, Zhu, Pan, Xue, Zhu, Chen, Li, Chen, Xie, Guo, and Xue]{SparkTTS}
Xinsheng Wang, Mingqi Jiang, Ziyang Ma, Ziyu Zhang, Songxiang Liu, Linqin Li, Zheng Liang, Qixi Zheng, Rui Wang, Xiaoqin Feng, Weizhen Bian, Zhen Ye, Sitong Cheng, Ruibin Yuan, Zhixian Zhao, Xinfa Zhu, Jiahao Pan, Liumeng Xue, Pengcheng Zhu, Yunlin Chen, Zhifei Li, Xie Chen, Lei Xie, Yike Guo, and Wei Xue.
\newblock {Spark-TTS}: {A}n efficient llm-based text-to-speech model with single-stream decoupled speech tokens.
\newblock \emph{arXiv preprint arXiv:2503.01710}, 2025{\natexlab{a}}.

\bibitem[Wang et~al.(2025{\natexlab{b}})Wang, Zheng, Zhang, Zhang, Liao, and Wu]{wang2025metis}
Yuancheng Wang, Jiachen Zheng, Junan Zhang, Xueyao Zhang, Huan Liao, and Zhizheng Wu.
\newblock Metis: A foundation speech generation model with masked generative pre-training.
\newblock \emph{arXiv preprint arXiv:2502.03128}, 2025{\natexlab{b}}.

\bibitem[Wang et~al.(2023{\natexlab{b}})Wang, Chen, Xie, Tian, and Wang]{wang2023lm}
Zhichao Wang, Yuanzhe Chen, Lei Xie, Qiao Tian, and Yuping Wang.
\newblock {LM-VC}: Zero-shot voice conversion via speech generation based on language models.
\newblock \emph{IEEE Signal Processing Letters}, 2023{\natexlab{b}}.

\bibitem[Wang et~al.(2024{\natexlab{b}})Wang, Zhu, Zhang, Lv, Jiang, Zhao, and Xie]{wang2024selm}
Ziqian Wang, Xinfa Zhu, Zihan Zhang, YuanJun Lv, Ning Jiang, Guoqing Zhao, and Lei Xie.
\newblock Selm: Speech enhancement using discrete tokens and language models.
\newblock In \emph{ICASSP 2024-2024 IEEE International Conference on Acoustics, Speech and Signal Processing (ICASSP)}, pp.\  11561--11565. IEEE, 2024{\natexlab{b}}.

\bibitem[Wang et~al.(2025{\natexlab{c}})Wang, Liu, Zhu, Li, Kang, Yao, Xia, Huang, and Xie]{wang2025uniflow}
Ziqian Wang, Zikai Liu, Yike Zhu, Xingchen Li, Boyi Kang, Jixun Yao, Xianjun Xia, Chuanzeng Huang, and Lei Xie.
\newblock {UniFlow}: Unifying speech front-end tasks via continuous generative modeling.
\newblock \emph{arXiv preprint arXiv:2508.07558}, 2025{\natexlab{c}}.
\newblock \doi{10.48550/arXiv.2508.07558}.
\newblock URL \url{https://arxiv.org/abs/2508.07558}.

\bibitem[Welker et~al.(2022)Welker, Richter, and Gerkmann]{welker22_interspeech}
Simon Welker, Julius Richter, and Timo Gerkmann.
\newblock Speech enhancement with score-based generative models in the complex {STFT} domain.
\newblock In \emph{Interspeech}, pp.\  2928--2932, 2022.
\newblock \doi{10.21437/Interspeech.2022-10653}.

\bibitem[Wichern et~al.(2019)Wichern, Antognini, Flynn, Zhu, McQuinn, Crow, Manilow, and Roux]{wichern2019wham}
Gordon Wichern, Joe Antognini, Michael Flynn, Licheng~Richard Zhu, Emmett McQuinn, Dwight Crow, Ethan Manilow, and Jonathan~Le Roux.
\newblock {WHAM!}: {E}xtending speech separation to noisy environments.
\newblock In \emph{Proceedings of the Conference of the International Speech Communication Association}, pp.\  1368--1372, 2019.

\bibitem[Williamson \& Wang(2017)Williamson and Wang]{tfmask}
Donald~S. Williamson and DeLiang Wang.
\newblock Time-frequency masking in the complex domain for speech dereverberation and denoising.
\newblock \emph{IEEE/ACM Transactions on Audio, Speech, and Language Processing}, 25\penalty0 (7):\penalty0 1492--1501, 2017.
\newblock \doi{10.1109/TASLP.2017.2696307}.

\bibitem[Wu et~al.(2023)Wu, Chen, Zhang, Hui, Berg-Kirkpatrick, and Dubnov]{clap}
Yusong Wu, Ke~Chen, Tianyu Zhang, Yuchen Hui, Taylor Berg-Kirkpatrick, and Shlomo Dubnov.
\newblock Large-scale contrastive language-audio pretraining with feature fusion and keyword-to-caption augmentation.
\newblock In \emph{IEEE International Conference on Acoustics, Speech and Signal Processing}, pp.\  1--5, 2023.
\newblock \doi{10.1109/ICASSP49357.2023.10095969}.

\bibitem[Xu et~al.(2025)Xu, Mei, Zheng, Tao, Xie, Zhang, Liu, Wu, Yan, Wu, Zhang, and Wu]{xu2025uniflowaudio}
Xuenan Xu, Jiahao Mei, Zihao Zheng, Ye~Tao, Zeyu Xie, Yaoyun Zhang, Haohe Liu, Yuning Wu, Ming Yan, Wen Wu, Chao Zhang, and Mengyue Wu.
\newblock {UniFlow-Audio}: Unified flow matching for audio generation from omni-modalities.
\newblock \emph{arXiv preprint arXiv:2509.24391}, 2025.

\bibitem[Yan et~al.(2025)Yan, Liu, Xue, Liang, and Xue]{unise}
Haoyin Yan, Chengwei Liu, Shaofei Xue, Xiaotao Liang, and Zheng Xue.
\newblock {UniSE}: A unified framework for decoder-only autoregressive lm-based speech enhancement.
\newblock \emph{arXiv preprint arXiv:2510.20441}, 2025.

\bibitem[Yang et~al.(2024)Yang, Tian, Tan, Huang, Liu, Guo, Chang, Shi, Bian, Zhao, et~al.]{yang2024uniaudio}
Dongchao Yang, Jinchuan Tian, Xu~Tan, Rongjie Huang, Songxiang Liu, Haohan Guo, Xuankai Chang, Jiatong Shi, Jiang Bian, Zhou Zhao, et~al.
\newblock {UniAudio}: {T}owards universal audio generation with large language models.
\newblock In \emph{Proceedings of the International Conference on Learning Representations}, 2024.

\bibitem[Yao et~al.(2025)Yao, Liu, Chen, Hu, Chng, and Xie]{yao2025gense}
Jixun Yao, Hexin Liu, Chen Chen, Yuchen Hu, EngSiong Chng, and Lei Xie.
\newblock Gen{SE}: Generative speech enhancement via language models using hierarchical modeling.
\newblock In \emph{The Thirteenth International Conference on Learning Representations}, 2025.

\bibitem[Ye et~al.(2024{\natexlab{a}})Ye, Sun, Lei, Lin, Tan, Dai, Kong, Chen, Pan, Liu, et~al.]{xcodec}
Zhen Ye, Peiwen Sun, Jiahe Lei, Hongzhan Lin, Xu~Tan, Zheqi Dai, Qiuqiang Kong, Jianyi Chen, Jiahao Pan, Qifeng Liu, et~al.
\newblock Codec does matter: Exploring the semantic shortcoming of codec for audio language model.
\newblock \emph{arXiv preprint arXiv:2408.17175}, 2024{\natexlab{a}}.

\bibitem[Ye et~al.(2024{\natexlab{b}})Ye, Sun, Lei, Lin, Tan, Dai, Kong, Chen, Pan, Liu, et~al.]{ye2024codec}
Zhen Ye, Peiwen Sun, Jiahe Lei, Hongzhan Lin, Xu~Tan, Zheqi Dai, Qiuqiang Kong, Jianyi Chen, Jiahao Pan, Qifeng Liu, et~al.
\newblock Codec does matter: Exploring the semantic shortcoming of codec for audio language model.
\newblock \emph{arXiv preprint arXiv:2408.17175}, 2024{\natexlab{b}}.

\bibitem[Ye et~al.(2025)Ye, Zhu, Chan, Wang, Tan, Lei, Peng, Liu, Jin, Dai, Lin, Chen, Du, Xue, Chen, Li, Xie, Kong, Guo, and Xue]{xcodec2}
Zhen Ye, Xinfa Zhu, Chi-Min Chan, Xinsheng Wang, Xu~Tan, Jiahe Lei, Yi~Peng, Haohe Liu, Yizhu Jin, Zheqi Dai, Hongzhan Lin, Jianyi Chen, Xingjian Du, Liumeng Xue, Yunlin Chen, Zhifei Li, Lei Xie, Qiuqiang Kong, Yike Guo, and Wei Xue.
\newblock Llasa: Scaling train-time and inference-time compute for llama-based speech synthesis.
\newblock \emph{arXiv preprint arXiv:2502.04128}, 2025.

\bibitem[Yuan et~al.(2025)Yuan, Liu, Liu, Plumbley, and Wang]{yuan2025flowsep}
Yi~Yuan, Xubo Liu, Haohe Liu, Mark~D Plumbley, and Wenwu Wang.
\newblock {FlowSep}: {L}anguage-queried sound separation with rectified flow matching.
\newblock In \emph{IEEE International Conference on Acoustics, Speech and Signal Processing}, pp.\  1--5. IEEE, 2025.

\bibitem[Zeghidour et~al.(2021)Zeghidour, Luebs, Omran, Skoglund, and Tagliasacchi]{soundstream}
Neil Zeghidour, Alejandro Luebs, Ahmed Omran, Jan Skoglund, and Marco Tagliasacchi.
\newblock Soundstream: An end-to-end neural audio codec.
\newblock \emph{IEEE/ACM Transactions on Audio, Speech, and Language Processing}, 30:\penalty0 495--507, 2021.

\bibitem[Zhang et~al.(2025)Zhang, Yang, Fang, Wang, Zhang, Wang, Fan, and Wu]{zhang2025anyenhance}
Junan Zhang, Jing Yang, Zihao Fang, Yuancheng Wang, Zehua Zhang, Zhuo Wang, Fan Fan, and Zhizheng Wu.
\newblock Anyenhance: A unified generative model with prompt-guidance and self-critic for voice enhancement.
\newblock \emph{IEEE Transactions on Audio, Speech and Language Processing}, 33:\penalty0 3085–3098, 2025.

\bibitem[Zhang et~al.(2024{\natexlab{a}})Zhang, Zhang, Peng, Tang, Manohar, Liu, Hwang, Li, Wang, Chan, Huang, Wu, and Ma]{vevo}
Xueyao Zhang, Xiaohui Zhang, Kainan Peng, Zhenyu Tang, Vimal Manohar, Yingru Liu, Jeff Hwang, Dangna Li, Yuhao Wang, Julian Chan, Yuan Huang, Zhizheng Wu, and Mingbo Ma.
\newblock Vevo: Controllable zero-shot voice imitation with self-supervised disentanglement.
\newblock \emph{OpenReview}, 2024{\natexlab{a}}.

\bibitem[Zhang et~al.(2024{\natexlab{b}})Zhang, Sun, Xia, Huang, Xiao, and Xie]{BS-PLCNet}
Zihan Zhang, Jiayao Sun, Xianjun Xia, Chuanzeng Huang, Yijian Xiao, and Lei Xie.
\newblock Bs-plcnet: Band-split packet loss concealment network with multi-task learning framework and multi-discriminators, 2024{\natexlab{b}}.
\newblock URL \url{https://arxiv.org/abs/2401.03687}.

\bibitem[Zhao et~al.(2022)Zhao, Ma, Watcharasupat, and Gan]{frcrn}
Shengkui Zhao, Bin Ma, Karn~N. Watcharasupat, and Woon-Seng Gan.
\newblock {FRCRN}: Boosting feature representation using frequency recurrence for monaural speech enhancement.
\newblock In \emph{Proc. ICASSP}, pp.\  9281--9285, 2022.

\bibitem[Zhao et~al.(2024)Zhao, Ma, Ni, Zhang, Wang, Nguyen, Zhou, Yip, Ng, and Ma]{MossFormer2}
Shengkui Zhao, Yukun Ma, Chongjia Ni, Chong Zhang, Hao Wang, Trung~Hieu Nguyen, Kun Zhou, Jia~Qi Yip, Dianwen Ng, and Bin Ma.
\newblock {MossFormer2}: Combining {Transformer} and {RNN}-free recurrent network for enhanced time-domain monaural speech separation.
\newblock In \emph{Proc. ICASSP}, pp.\  10356--10360, 2024.

\end{thebibliography}
